\documentclass[12pt]{article}

\usepackage{a4}\usepackage{epsf}\usepackage{latexsym}

\usepackage{Abbreviations}

\begin{document}

\title{Temperature phase transition and an effective expansion
parameter in the O(N)-model} 
\author{ {\sc M. Bordag}\thanks{e-mail: Michael.Bordag@itp.uni-leipzig.de} \\ 
\small University of Leipzig, Institute for Theoretical Physics\\ 
\small Augustusplatz 10/11, 04109 Leipzig, Germany\\ 
\small and\\ 
{\sc V. Skalozub}\thanks{e-mail: Skalozub@ff.dsu.dp.ua}\\ 
\small Dnepropetrovsk National University,  49050 Dnepropetrovsk, Ukraine} 
\maketitle
\begin{abstract}
The temperature phase transition in the $N$-component scalar field
 theory with spontaneous symmetry breaking is investigated in the
 perturbative approach. The second Legendre transform is used together
 with the consideration of the gap equations in the extrema of the
 free energy.  Resummations are performed on the super daisy level and
 beyond.  The phase transition turns out to be  weakly of first
 order. The diagrams beyond the super daisy ones which are calculated
 correspond to next-to-next-to-leading order in $1/N$. It is shown
 that these diagrams do not alter the phase transition
 qualitatively. In the limit $N$ goes to infinity the phase transition
 becomes second order. A comparison with other approaches is done.
\end{abstract}
\thispagestyle{empty}


\section{Introduction}

The temperature induced phase transition in the N-component scalar
Field theory (O(N)-model) has been investigated by many authors.  The
interest to this problem results, in particular, from the importance
of the spontaneous symmetry breaking and symmetry restoration at
finite temperature in various modern field theoretical models of
elementary particles including different content of scalar fields.
The present status of this problem is characterized by a general
opinion that the phase transition in the O(N)-models (including
 N=1) is of second order (see, for instance, the standard text books
\cite{zinnjustin,linde1,Kapusta}.) This conclusion results mainly from
the non perturbative methods - lattice simulations, average action
(flow equation) methods, large-N expansion  \cite{Tetradis:1993xd,Reuter:1993rm,
Adams:1995cv,Montvay,Fodor1995af}. In opposite, a first order phase transition was found in the most perturbative approaches \cite{Takahashi1985a,Carrington:1992hz,Arnold1992a}. Here, the key problem 
- the necessity of resummation of the infrared divergences of a series in
 the coupling constant - has been overcome by applying the super daisy
 resummation.  However, after that another problem - the lack of the
 smallness of the effective expansion parameter near the phase transition 
- remains. This observation prevents conclusive results from the resummed perturbation theory and it leaves room for the expectation that a further
 resummation of the perturbation expansion is able to weaken the first order
 character of the expansion making it a second order at the end.  In recent
 papers \cite{Ogure:1998je,Ogure:1998xu} by an auxiliary-mass method it was
found that the phase transition in either the $O(1)$- or the $O(N)$-model is 
of second order. Moreover, it was also stated in Ref. \cite{Ogure:1998je}
 that in perturbation theory at the two-loop level the phase transition is also of second order when diagrams beyond the super daisy type are taken into account.  Because of the importance of these results we shall discuss them in more detail in what follows. Here, we note that the effective potential calculated by this method has an imaginary part in its minimum which means an
inconsistency of the calculation procedure adopted.

In our previous paper \cite{Bordag:2000tb} within the $O(1)$-model we developed as a new method the combination of the gap equation resulting from the second Legendre transform with the condition of the free energy considered as a  function of the condensate to be in an extremum. This allows for the exclusion of the condensate value from the gap equation and results in considerable simplifications. For instance, in the super daisy approximation  we obtained very simple explicit formulas showing clearly a first order phase transition. We
applied our method to a certain class of graphs beyond this approximation and obtained an increase of the first order character of the transition.

In the present paper we generalize our method to the $O(N)$-model. Again, in the extrema of the free energy, we exclude the condensate from the gap equations which form a set of two equations for the Higgs and Goldstone field masses. We consider in detail both, the super daisy resummation and the next class of graphs beyond.    We find that the phase transition is of first order for any finite $N$ turning into a second order one in the limit of $N$ goes to infinity. At the same time, the observed phase transition is very close to a second order one and differs from it in the next-to-leading order in $1/N$ for finite but large $N$. We demonstrate the effectiveness of our method by calculating a certain class of graphs beyond the super daisy approximation in the $O(N)$-model, namely those which correspond to the next-to-next-to-leading order for large $N$. Here we are left with a mostly  numerical approach to the gap
equations confirming the qualitative stability of our previous results. The diagram mentioned in Ref. \cite{Ogure:1998je}  is also included in the carried out resummations.  Its influence is discussed in the last section.

The paper is organized as follows. In the next section we collect the basic formulas of the resummation method based on the second Legendre transform together with our method to consider the gap equations in the extrema of the free energy. In the third section we consider the case corresponding to the super daisy approximation.  In the fourth section the summation of the bubble chain diagrams is carried out. Conclusions are drawn in the last section. Some technical material is banned into the appendix.

\section{Basic formulas}

We consider a scalar $N$-component
 field theory in $(3+1)$ dimensions
in the Euclidean space-time. The action
 reads
\bea\label{action}
S[\phi]&=&\int \d x \left(
\frac12  
 \phi(x){\bf K}\phi(x)-\frac{\la}{4N}(\phi(x))^{4}    \right) ,
\eea
where ${\bf K}=\Box -m^{2}$ is the kernel of the free action.
 The
sign is chosen so that the vacuum Green functions are given by
 the
functional representation
\be\label{Z}
Z\equiv e^{W}=\int D\phi
 ~{e^{S}}\,,
\ee
where $W$ is the functional of the connected Green
 functions. It is
connected by the relation 
\be\label{W} W=-FT
\ee

with the free energy $F$. We consider the theory at finite temperature $T$
in the Matsubara formalism so that the loop integrations are given by
\be\label{Tr}
Tr_p=T\sum_{l=-\infty}^{\infty}\int {\d^{3}\vec{p}\over
 (2\pi)^{3}}
\ee
with the momentum $p=(2\pi T l, \vec{p})$ ($l\in Z$).
  The symbol
$\phi$ is here a condensed notation for $N$ scalar fields,
$\phi=\{\phi_1,\phi_2,\dots\}$. In the action Eq.\Ref{action} the
following summation over the internal indices is assumed:
$\phi^2=
\sum_{a=1}^N\phi_a^2$ and
$\phi^4=\left(\sum_{a=1}^N\phi_a^2\right)^2$. 
 In the present paper, we perform a standard renormalization procedure at
 zero temperature.

The second Legendre transform is introduced as follows. Consider as an arbitrary functional argument the full propagator $\beta$, instead of being the free one
\be\label{freeprop}\Delta=-K^{-1},
\ee
Then the functional $W$ is a functional depending on $\beta$: $W[\beta]$. Here and in the following we denote a functional dependence by squared brackets in opposite to the dependence of a function on its argument. So, for example, in momentum representation, $\beta$ is a function of the momentum $p$: $\beta(p)$,
 Whereas $W[\beta]$ is not, of course.  In this sense the action Eq.\Ref{action}
is also a functional of $\phi$: $S[\phi]$.

Now, the second Legendre transform results in the representation
\be\label{Wallg}
W=S[0]+\frac12 Tr \log \beta -\frac12 Tr \Delta^{-1}\beta +
W_{2}[\beta]
\ee
of the connected Green functions, where $\beta$ is subjected to the
Schwinger-Dyson (SD) equation
\be\label{SDallg}
\beta^{-1}(p)=\Delta^{-1}-\Sigma[\beta](p) \,.
\ee
Here, $\Sigma[\beta](p)$ is the functional of all self energy graphs
with no propagator insertions,
\be\label{Sigallg}
\Sigma[\beta](p)=2{\delta
 W_{2}\over \delta \beta(p)} ,
\ee
where $\delta$ is the functional  derivative, see, e.g., \cite{2PI} for
details. $W_{2}[\beta]$ is the sum of all two particle irreducible
(2PI) graphs taken out of the connected Green functions,
\be\label{W22PI} W_2[\beta]=\sum_{\rm 2PI-graphs}
W[\beta] 
\ee
with the function $\beta(p)$ on the lines.

The above formulas are written in condensed notations. So, they are valid for any theory with any number of fields. Below, we will write down them explicitly
in the $O(N)$-model. But before doing that we introduce the spontaneous symmetry breaking by turning the sign of the mass term in the free propagator, $m^2\to -m^2$ in $\Delta$ so that it reads now $\Delta=p^2-m^2$ in the momentum representation.  Then, on the tree level, the minimum of the energy is on the shifted fields, $\phi_1\to\eta+v$, where $v$ is the field condensate.  After that we change the notations according to $\{\phi_1, \phi_2,\dots\}\to
\{\eta+v,\phi_1,\phi_2,\dots\}$ where $\eta$ is the Higgs  field and
$\{\phi_1,\phi_2,\dots\} \ \ (a=1,2,\dots,N-1)$ are the Goldstone
fields which are symmetric under the residual $O(N-1)$ symmetry.

In terms of the new variables the action reads
\bea\label{Sef} S[\eta,\Phi]&=&\int
 dx \left\{
\frac{m^2}{2}v^2-\frac{\la}{4N}v^4 +\eta(m^2-\frac{\la}{N}v^2)v
\right. \nn\\ &&~~~~~~ +\frac12 \eta(\Box-\mu_\eta^2)\eta +\frac12
\phi(
\Box-\mu_ \phi^2)\phi \nn\\ && ~~~~~~ \left. -\frac{\la}{4N}
\left( 
\eta^4+4\eta^3 v +2\left(\eta^2+2\eta
v\right)\phi^2+\phi^4\right) \right\}
\eea
 with $\mu_\eta=-m^2+3\frac{\la}{N}v^2$ and
$\mu_\phi=-m^2+
\frac{\la}{N}v^2$. Again, the summation over the internal indices is assumed.
 In momentum representation, the corresponding free propagators are
\be
\label{freep} 
\Delta_\eta=p^2+\mu_\eta^2  \qquad \mbox{\rm and} \qquad 
\Delta_\phi=p^2+\mu_\phi^2.
\ee
The propagator for the Goldstone fields is
 diagonal,
$(\Delta_\phi)_{ab}=\delta_{ab} \ \Delta_{\phi}$. With these
 notations, the representation \Ref{Wallg} for the connected vacuum Green
functions takes the form
\bea\label{WON}
W&=&
S[0]+\frac12 Tr 
\log \beta_\eta +\frac{N-1}{2} Tr \log \beta_\phi \nn \\
&&-\frac12 Tr
 \Delta_\eta^{-1}\beta_\eta
-\frac{N-1}2 Tr \Delta_\phi^{-1}\beta_\phi 
+W_{2}[\beta_\eta,\beta_\phi],
\eea
where $W_{2}[\beta_\eta,\beta_\phi]$ is the sum of all 2PI graphs with
the propagators $\beta_\eta$ and $\beta_\phi$ on the lines corresponding to the fields $\eta$ and $\phi$.  In the $O(N)$-model, the SD-equations form a set of $N$ equations which are equivalent to one equation due to the symmetry. In the considered case of the broken symmetry, two equations remain:
\bea\label{SD2}
\beta^{-1}_\eta(p)&=&\Delta^{-1}_\eta-\Sigma_\eta[\beta_\eta,
\beta_\phi](p)
, \nn\\
\beta^{-1}_\phi(p)&=&\Delta^{-1}_\phi-
\Sigma_\phi[\beta_\eta,\beta_\phi](p)
, \eea
where following \Ref{Sigallg} the self energy functionals are given by
\bea \label{SigON}
\Sigma_
\eta[\beta_\eta,\beta_\phi](p)&=&2\frac{\delta
W_{2}[\beta_\eta,
\beta_\phi]}{\delta\beta_\eta(p)}, \nn \\
\Sigma_\phi[
\beta_\eta,\beta_\phi](p)&=&\frac{2}{N-1} \frac{\delta
W_{2}[
\beta_\eta,\beta_\phi]}{\delta\phi_\eta(p)}.  \eea
Here, the simple relation
$(\Sigma_\phi)_{ab}=\delta_{ab}\Sigma_\phi=2\frac{\delta 
W_2}{\delta
(\beta_\phi)_{ab}}=\delta_{ab}\frac{2}{N-1}\frac{\delta W_2}
{\delta
\beta_\phi}$ was taken into account.

At sufficiently high temperature the symmetry is restored. This means
that the value of the condensate is zero, $v=0$, and the masses of the
Higgs and the Goldstone fields are equal, $M_\eta=M_\phi$. Then the
SD-equations \Ref{SD2} reduce to one equation, Eq. \Ref{SDallg}. At low
temperature, the symmetry is broken and $v$ is non zero. The masses are
different and we have to consider two equations, Eqs. \Ref{SD2}. Then, the free energy is the function of the condensate $v$. It has a minimum at $v>0$ which can be seen already on the tree level. It may have more extrema. In that case 
due to the continuity of the free energy as a function of $v$ an additional
extremum for $v>0$ must be a maximum indicating a first order phase transition.

Following \cite{Bordag:2000tb}, we consider the free energy in its extrema. We take the derivative of $W$ with respect to the condensate $v^2$,
\be\label{null}
\frac{d }{d v^2}
 W_2[\beta_\eta,\beta_\phi]
=
\frac{m^2}{2}-\frac{\la}{2N}v^2-
\frac{3\la}{2N} \Tr\beta_\eta
-\frac{\la(N-1)}{2N}\Tr\beta_\phi
+
\frac{\pa W_2[\beta_\eta,\beta_\phi]}{\pa v^2}.
\ee
In general, in this expression there are contributions proportional to
${\pa \beta_\eta}/{\pa
 v^2}$ and ${\pa\beta_\phi}/{\pa v^2}$. But
they vanish by means of the SD-equations \Ref{SD2}. In Eq. \Ref{null},
$\frac{\pa W_2}{\pa v^2}$ denotes the derivative with respect to the explicit dependence on $v$ entering through the vertex factors.

Demanding now the derivative given by Eq. \Ref{null} to vanish,
\be\label{nulll}\frac{d W_2[\beta_\eta,\beta\phi]}
{d v^2}=0,
\ee
we obtain the following equation,
\be\label{v}
\frac{
\la}{N}v^2=m^2-3\frac{\la}{N}\Tr\beta_\eta 
-\frac{\la(N-1)}{N}\Tr
\beta_\phi +2\frac{\pa W_2}{\pa v^2},
\ee
which has to be considered together with the SD-equations \Ref{SD2}. 

In order to proceed it is necessary to make approximations. First of
all, we put the momentum $p$ in the SD-equations equal to zero, $p=0$,
 that is, we turn to the corresponding gap equations. This is reasonable 
because we are interested in  the infrared behavior and the gap in the spectrum is the decisive quantity. So we make the ansatz
\bea\label{betam} 
 \beta_\eta &=& p^2+M_\eta^2\nn \\
 \beta_\phi &=& p^2+M_\phi^2
\eea
for the propagators and consider the SD equations \Ref{SD2} at $p=0$.
The assumption connected with this approximation is that the \uv
behavior of the propagator is mostly the same for the full solution
and for \Ref{betam}, that the behavior for mediate momenta is not
essential and that the important information is contained in the gap,
i.e., in $M_\eta$ and in $M_\phi$.  We note that in the super daisy
approximation the corresponding contribution to $\Sigma$ is
independent of $p$ and the approximation \Ref{betam} becomes an exact
relation.

The next approximation concerns the functional $W_2$. Here, we restrict
ourselves to graphs with at most two insertions of the trilinear vertex. In
analytic notation we have
\bea\label{W2app0}
W_2[
\beta_\eta,\beta_\phi]&=&W_2^{\rm SD}+D+v^2\Gamma+O(v^4).
\eea
Contributions proportional to $v^4$ and higher powers of $v^2$ are dropped.
 Furthermore, we wrote down explicitely the lowest graphs, i.e., the graphs
with one vertex, and introduced the notation $D$ for the remaining ones. By construction, the graphs in $D$ do not contain $v$-dependent vertices and
 have at least two vertices.  The graphs with one vertex are \vspace{-1.8cm}
\bea\label{W2SD}W_2^{\rm SD}&=&\frac18
\hspace{-0.5cm}
\epsfxsize=3.7cm\raisebox{-2.6cm}{\epsffile{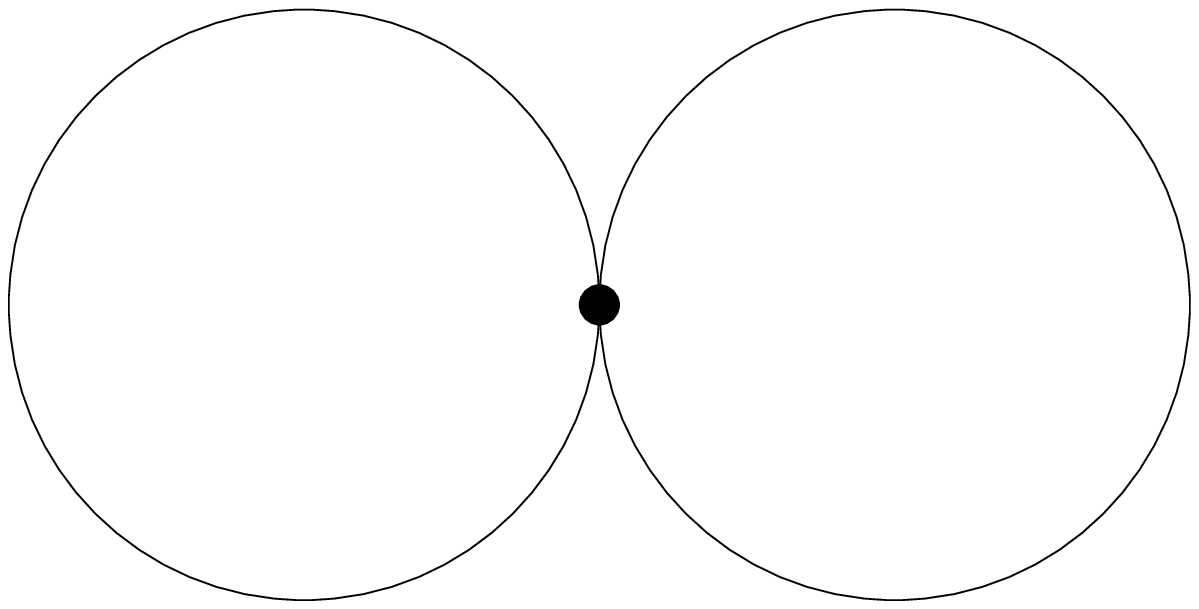}}
\hspace
{-0.5cm}
+
\frac14
\hspace{-0.5cm}
\epsfxsize=3.7cm\raisebox{-2.6cm}
{\epsffile{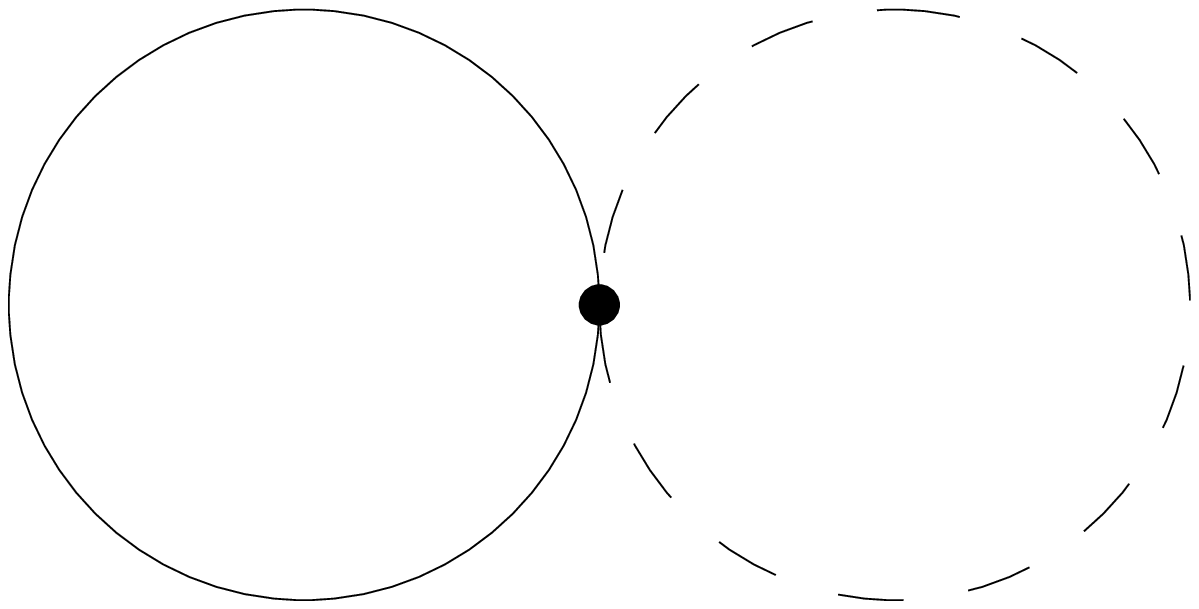}} 
\hspace{-0.5cm}
+
\frac14
\hspace{-0.5cm}
\epsfxsize=3.7cm\raisebox{-2.6cm}{\epsffile{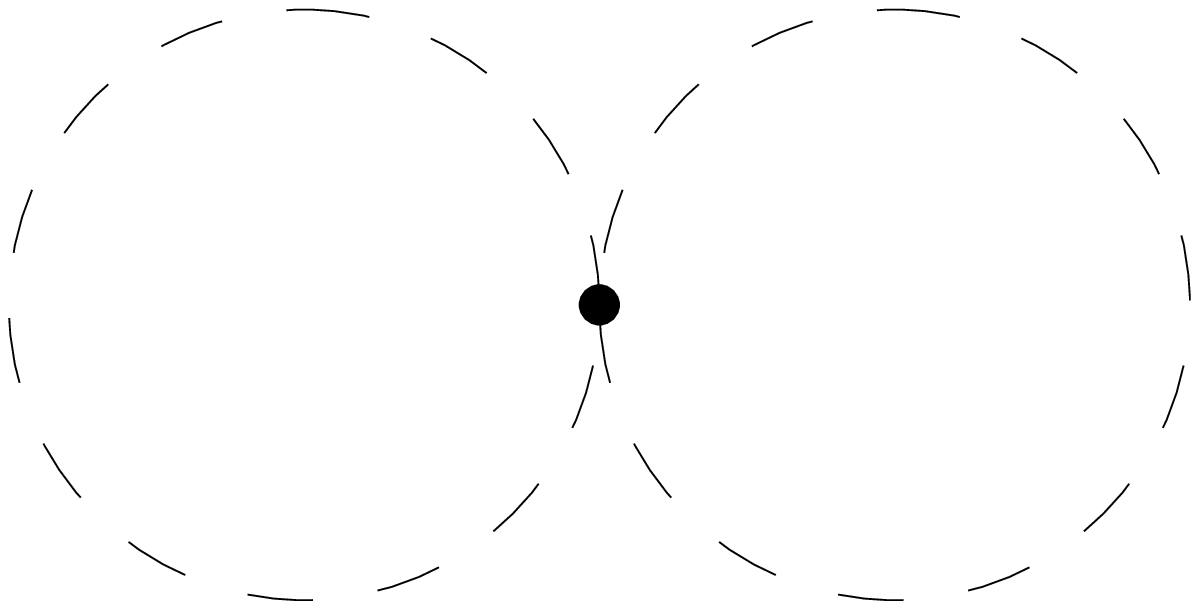}} 
\nn \\
[-1.5cm]
&=&-\frac{3\la}{4N}(\Sigma_\eta^{(0)})^2 
-\frac{\la}{2}
\frac{N-1}{N}\Sigma_\eta^{(0)} \ \Sigma_\phi^{(0)}
-\frac{\la}{4}
\frac{N^2-1}{N}(\Sigma_\phi^{(0)})^2,
\eea
where the solid line is the Higgs field propagator, $\beta_\eta$, and the
broken line is the Goldstone one, $\beta_\phi$. Taken as the approximation to $W_2$ these graphs generate all super daisy graphs. The vertex factors are adduced in the Appendix, Table 2. In $W_2^{\rm SD}$ Eq. \Ref{W2SD} we introduced the notations \vspace{-1.8cm}
\bea\label{dfs} \Sigma_\eta^{(0)} &=& \Tr \beta_\eta 
= 
\hspace{-0.5cm}
\epsfxsize=3.7cm\raisebox{-2.6cm}{\epsffile{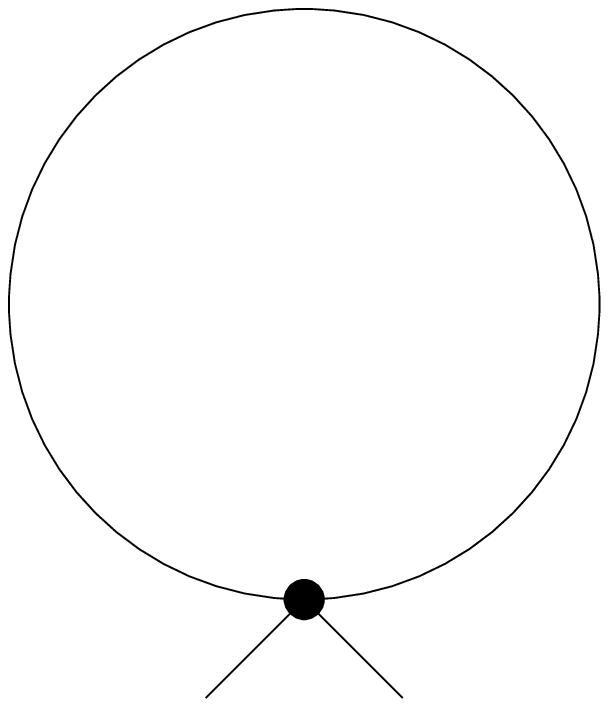}}
\hspace{-0.5cm},
 \nn \\[-3.5cm]
\Sigma_\phi^{(0)} &=& \Tr \beta_\phi =   
\hspace{-0.5cm}
\epsfxsize=3.7cm\raisebox{-2.6cm}{\epsffile{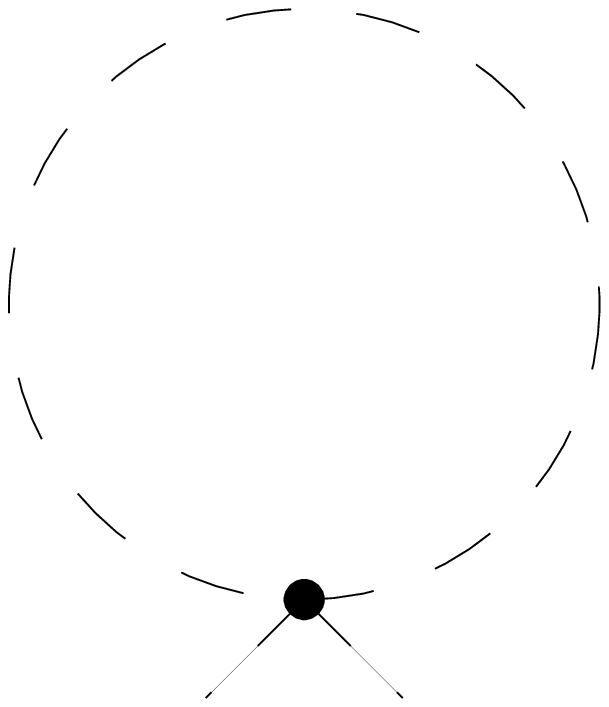}}
\hspace{-0.5cm}.
\eea
\vspace{-2cm}

Now, let us rewrite the gap equations in this approximation. In the restored 
phase, we have to put $v=0$ and to take the corresponding functional 
derivatives.  We obtain
\bea\label{geq1r}
 M_\eta^2&=&-m^2+\frac{3\la}
{N}\Sigma_\eta^{(0)}
+\la\frac{N-1}{N}\Sigma_\phi^{(0)}-2\frac{\delta D}
{\delta \beta_\eta}, \nn \\
 M_\phi^2&=&-m^2+\frac{\la}{N}\Sigma_\eta^{(0)}
+\la\frac{N+1}{N}\Sigma_\phi^{(0)}
-\frac{2}{N-1}\frac{\delta D}{\delta 
\beta_\phi} .
\eea
Taking into account that $\frac{\delta D}{\delta
\beta_
\eta}=\frac{1}{N-1}\frac{\delta D}{\delta \beta_\phi}$ holds at $M_\eta=
M_\phi$  the two equations become equal for symmetry reasons and we are left with one equation for the mass in the restored phase which we denote as $M_r$.

In the broken phase, in the extremum of the free energy the condensate is given by Eq. \Ref{v}. First, using the approximation \Ref{W2app0} we rewrite the gap equations \Ref{SD2} as follows
\bea\label
{geq1a} 
 M_\eta^2&=&-m^2+\frac{3\la}{N}v^2
+\frac{3\la}{N}\Sigma_\eta^{(0)}%
 \\ && ~~~~~~~~~~~~~~
+\la\frac{N-1}{N}\Sigma_\phi^{(0)}
-2v^2\frac{\delta 
\Gamma}{\delta \beta_\eta}
-2\frac{\delta D}{\delta \beta_\eta},
   \\
\label
{geq1b} 
 M_\phi^2&=&-m^2+\frac{\la}{N}v^2
+\frac{\la}{N}\Sigma_\eta^{(0)} %
 \\ && ~~~~~~~~~~~~~~
+\la\frac{N+1}{N}\Sigma_\phi^{(0)}
-v^2\frac{2}{N-1}
\frac{\delta \Gamma}{\delta \beta_\eta}
-\frac{2}{N-1}\frac{\delta D}{\delta
 \beta_\phi}.  
\eea

The equation \Ref{v} resulting from the condition of the extremum takes the form
\be\label{exc}
\frac{\la v^2}{N}=m^2-\frac{
3\la}{N}\Sigma_\eta^{(0)}
-\la\frac{N-1}{N}\Sigma_\phi^{(0)}+2\Gamma .
\ee
Now, by using Eq. \Ref{exc} we rewrite the first gap equation, Eq.\Ref{geq1a}, in the form
\be\label{ge2a}M_\eta^2=
2\left(1-
\frac{N}{\la}\frac{\delta\Gamma}{\delta\beta_\eta}\right)
\left(
m^2-\frac{3\la}{N}\Sigma_\eta^{(0)}
-\la\frac{N-1}{N}\Sigma_\phi^{(0)}
+2\Gamma\right),
\ee
where we took  the relation $\Gamma=\frac{\delta D}{\delta\beta_\eta}$ into account. Then, by means of Eq. \Ref{geq1a}, we rewrite Eq. \Ref{exc} as
\be\label{Mv}\frac{\la}{N}v^2=
\left(1-\frac{N}{\la}
\frac{\delta\Gamma}{\delta\beta_\eta}\right)^{-1}
\frac{M_\eta^2}{2}.
\ee
The second gap equation can be rewritten in a similar way. First we insert
$v$ from Eq. \Ref{v} in its first appearance and from Eq. \Ref{Mv} in its second one. We obtain finally
\be\label{ge2b}
M_\phi^2=\frac{2\la}
{N}
\left(\Sigma_\phi^{(0)}
-\Sigma_\eta^{(0)}\right)
-\frac{M_\eta^2}
{N-1}
\frac{\frac{\delta\Gamma}{\delta\beta_\phi}}{\frac{\la}{N}
-\frac
{\delta\Gamma}{\delta\beta_\eta}}
+2\left(\Gamma-\frac{1}{N-1}\frac{
\delta D}{\delta\beta_\phi}\right) .
\ee
In this way all necessary general formulas are collected. In the next sections we will proceed with more specific approximations.


\section{Super daisy approximation (SDA)}

The super daisy approximation (sometimes the corresponding diagrams
are called also cactus or foam diagrams) is the result of summing up
all graphs containing lines closed up over one vertex. It is known that this
 resummation is equivalent to take in the functional $W_2$, Eq. \Ref{W22PI}, the
contribution of $W_2^{\rm SD}$, \Ref{W2SD}, only. So, we have to put in
the formulas given in the preceding section $\Gamma=D=0$.

First, we write down the functional $W$, Eq. \Ref{WON}, in the chosen
approximation. To that end we rewrite the contributions like $\Tr
\Delta^{-1}_\eta\beta_\eta$ by using the SD equations \Ref{SD2} and 
$\Sigma_\eta^{(0)}$ resp. $\Sigma_\eta^{(0)}$ in the right handsides. 
Multiplying by $\beta_\eta(p)$ resp. $\beta_\phi(p)$ and taking the trace,
 by means of Eq. \Ref{dfs} we have
\beao 
\Tr \Delta^{-1}_\eta 
\beta_\eta(p) = \left(\Sigma_\eta^{(0)}\right)^2 +c, \\
\Tr 
\Delta^{-1}_\phi \beta_\phi(p) = \left(\Sigma_\phi^{(0)}\right)^2 
+c,
\eeao
where $c$ is an irrelevant constant. Taking into account these relations we
 rewrite $W$ as
\be\label{Wsd} 
W(v^2,M_\eta,M_\phi)= \frac{m^2}{2}v^2
-\frac{\la}{N}v^4
+\frac12 \Tr\ln\beta_\eta+\frac{N-1}{2}\Tr\ln\beta_\phi
 -W_2^{\rm SD}
\ee
with $W_2^{\rm SD}$ given by Eq. \Ref{W2SD}. Here we explictely denoted the
arguments $W$ dependes on.  

Now we turn to the gap equations. In the restored phase, we have $v=0$ and the masses are equal, $M_\eta=M_\phi\equiv M_r$.  There is only one equation following from Eqs. \Ref{geq1r}. Now it reads
\be\label{geq2r} M_r^2=-m^2+\la\frac{N+2}{N}
\Sigma^{(0)},
\ee
where in $\Sigma^{(0)}$ the propagator $\beta=1/(p^2
+M_r^2)$ has to be inserted, similar to Eqs. \Ref{dfs}.  The corresponding
 value of $W$ is $W(0,M_r,M_r)$.

In the broken phase, we first note Eq. \Ref{Mv} which takes the form
\be\label{Mvsd}   \frac{\la}{N}v^2=
\frac12 M_\eta^2 .
\ee
The gap equations \Ref{ge2a}, \Ref{ge2b} turn into
\bea\label{ge2} M_\eta^2&=& 2\left(m^2-\frac{3\la}{N} 
\Sigma_\eta^{(0)} 
-\la\frac{N-1}{N}\Sigma_\phi^{(0)}\right) ,
\nn\\
 M_\phi^2&=& \frac{2\la}{N} \left(\Sigma_\phi^{(0)} -\Sigma_
\eta^{(0)}\right) .
\eea

Finally, we take the high temperature expansion for the functions
entering. They are well known and we use them in the form given in our
previous papers \cite{bordagskalozub99,Bordag:2000tb}:
\bea\label{HighTa} V_1(M)\equiv -\frac12\Tr\ln\beta &=&
\frac{-\pi^2T^4}{90}+\frac{M^2T^2}{24}-\frac{M^3T}{12\pi}+\dots, \nn
 \\
\Sigma^{(0)}&=&\frac{T^2}{12}-\frac{MT}{4\pi}+\dots ,
\eea
where the corresponding masses $M_r$, $M_\eta$ or $M_\phi$ have to be
inserted. In Eqs. \Ref{HighTa}, the dots denote contributions suppressed by higher powers of $M/T$. 

In this approximation the gap equations read, in the restored phase,
\bea\label{gesdr}
M_r^2&=& -m^2+\frac{\la(N+2)T^2}{12N}
-2M_r\frac{\la(N+2)T}{8\pi N} ,
\eea
and in the broken phase,
\bea\label
{gesdb}
M_\eta^2&=& 2m^2-\frac{\la(N+2)T^2}{6N}+
2M_\eta\frac{3\la T}{4\pi
 N}+M_\phi\frac{\la(N-1)T}{2\pi N}, \nn \\
M_\phi^2&=& \frac{\la T}{2\pi N}
\left(M_\eta-M_\phi\right).
\eea
In the restored phase, Eq. \Ref{gesdr} can be solved simply with the result
\be\label{Mrsd} M_r=-\frac{\la(N+2)T}
{8\pi N}
+\sqrt{\left(\frac{\la(N+2)T}{8\pi N}\right)^2-m^2+
\frac{\la (N+2)
 T^2}{12N}} \,.
\ee
We observe that it has a single solution for positive mass as long as 
$T>T_-$, where
\be\label{T-sd} T_-=\sqrt{\frac{12N}{
\la(N+2)}}m
\ee
is the lower spinodal temperature. 

In the broken phase, the two gap equations \Ref{gesdb} form a system
of algebraic equations. While in the one component case there had been
one equation of second order (a quadratic equation like Eq. \Ref{gesdr}
in the restored phase), here the order is higher because there are two
coupled equations.  We proceed as follows. The second equation can be
solved with respect to $M_\phi$,
\be
\label{Mf} M_\phi(M_\eta)=-\frac{\la T}{4\pi N}
+\sqrt{\left(\frac{\la T}
{4\pi N}\right)^2+\frac{\la T}{2\pi N}M_\eta}.
\ee
This is a unique solution $M_\phi(M_\eta)$ for all $T$. Note the
special value $M_\phi(0)=0$ independent of $T$.

Now we insert $M_\phi(M_\eta)$ from \Ref{Mf} into the first equation in formula \Ref{gesdb},
\bea\label{gesdbi}
M_\eta^2&=& 2m^2
-\frac{\la(N+2)T^2}{6N}+
2M_\eta\frac{3\la T}{4\pi N}+\frac{\la(N-1)T}{2\pi
 N}M_\phi(M_\eta).
\eea
This equation can be written down as a fourth order algebraic equation.
But for a moment let us rewrite it  formally by solving as a quadratic equation for $M_\eta$ keeping $M_\phi(M_\eta)$,
\be\label{2eqf} 
M_\eta=\frac{3\la T}{4\pi N}\pm\sqrt{\left(\frac{3\la T}
{4\pi N}\right)^2
+2m^2-\frac{\la(N+2)T^2}{6N}+\frac{\la(N-1)T}{2\pi N}M_
\Phi(M_\eta)}.
\ee
For $T=0$ we have a solution for the upper sign which is clearly the
mass in the minimum of the free energy. Hence the solution with the
lower sign describes a maximum. This mass vanishes for $T=T_-$ (taking
into account Eqs. \Ref{T-sd} and \Ref{Mf}). 

\begin{figure}[h]\unitlength=1cm
\begin{picture}(5,7.5)
\put(0,-1){
 
\epsffile{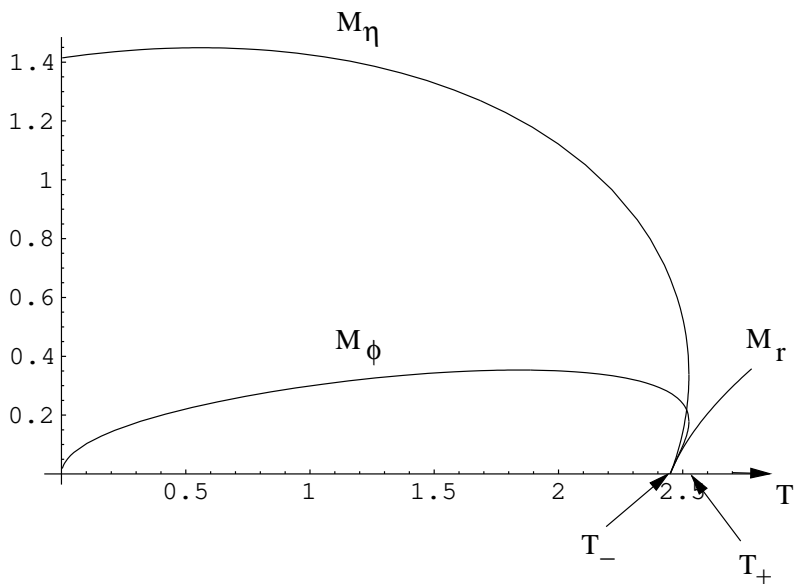}}
\end{picture}
\caption{The masses $M_\eta$, $M_\phi$ and $M_r$ as functions of the
temperature $T$ for $\la=1$, $N=2$. The upper parts of the curves for
$M_\eta$ and $M_\phi$ show the corresponding masses in the minimum of
the free energy, the lower parts show the masses in the maximum.}
\label{figure1}
\end
{figure}

Eq. \Ref{gesdbi} has an explicit solution, which is, however, too
large to be displayed here. It can be easily plotted and is shown in
Figure \ref{figure1}. We have chosen $\la =1$ in order to make the details near the phase transition better visible. But we can obtain the basic properties of this solution in the follwing way.  Assume we plot the rhs. versus lhs. of Eq. \Ref{gesdbi}. We observe that for small $T$ there is one solution. At $T=T_-$ a second solution appears and at some temperature $T_+$ both solutions merge and disappear. So $T_+$ has to be interpreted as the upper spinodal temperature.  In
this way we observe a first order phase transition.  Once we have an algebraic solution of the gap equations, we can give an algebraic expression for $T_+$. Again, it is too complicated to be displayed. However, for small $\la$, $T_+$  can be expressed as
\be\label{T+la}\frac{T_+}{T_-}=1+t_N \la+O(\la^2),
\ee
where the numbers $t_N$ are algebraic expressions in $N$. Some numerical
 values are given in Table 1. For $N=1$ we reobtain with $t_1=\frac{9}{
16\pi^2}$ the known result from the one component case in
 \cite{Bordag:2000tb}. In the limit of large $N$ we obtain
\be\label{tlargeN}t_N\sim\frac{9\la}{16\pi^2}\frac{1}{(2N)^{\frac23}}
\ee
which shows that the phase transition becomes second order for large $N$.


\begin{table}
\begin{tabular}{|r|r|r|r|r|r|r|r|}
\hline
 
$N$&0&1&2&3&4&5 \\ 
\hline
$t_N$&0.0569932&0.029511&0.0203949&0.0158314&0.0130797&0.0112318\\
 \hline 
\end{tabular}
\caption{Some numerical values of the coefficients
 $t_N$ in Eq. \Ref{T+la}}
\end{table}


\begin{figure}[h]\unitlength=1cm
\begin{picture}(5,6)
\put(0,-2.5){ 
\epsffile{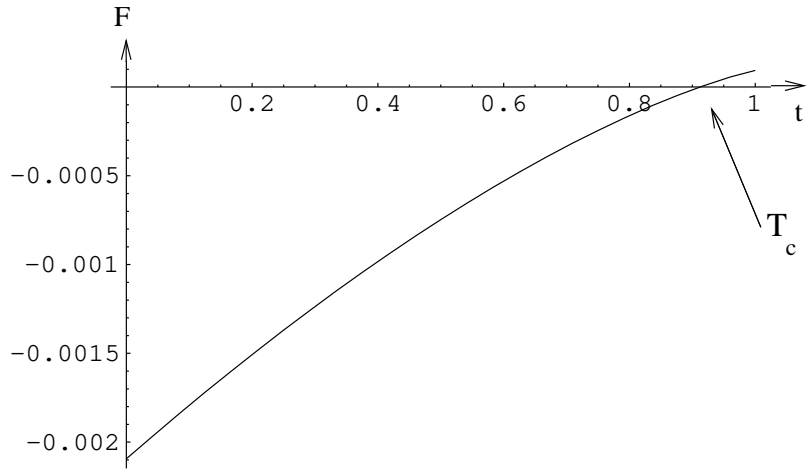}}
\end{picture}
\caption{The free energy $F$ as function of $t$ parameterizing by
 means
of $T=T_-+t(T_+-T_-)$ the temperature in betweeen $T_-$ and $T_+$.}
\label{figure2}
\end{figure}

Now, having solved the gap equation, we consider the functional $W$. Let $W_b$ be its value in the minimum (broken phase), given by Eq. \Ref{Wsd} with the masses inserted from the solutions discussed above and the condensate $v$ from Eq. \Ref{Mvsd}. Let $W_r=W(0,M_r,M_r)$ be the corresponding value in the restored phase. Again, the analytic expressions are too large to be displayed. So, we restrict ourselves to some numerical values. For example, for $\la=1$ and
 $N=2$ we have $T_-=2.44949 $ and $T_+=2.52514 $. In $T=T_-$ we have 
$W_b=-8.15081$ and $W_r=-8.14568$ (note that the free
energy differs in sign from $W$, see Eq. \Ref{W}). In $T=T_+$ the corresponding values are 
$W_b=-8.15081$ and $W_r=-9.1671$.  In
Figure \ref{figure2} we plotted 
$F=-\Delta W/T$ as function of the
temperature in the region of the phase
 transition, i.e., for $T_-\le
T\le T_+$.
The
temperature
 $T_c$ of the phase transition follows from $\Delta W\equiv
W_b-W_r=0$ to
 be $T_c=7.767$.


\section{Beyond the super daisy approximation (BSDA)}

In this section we investigate the phase transition beyond the SDA. In the chosen approach of representing the free energy by the 2PI functional $W_2$ an approximation is given by chosing an approximation for $W_2$.  While we in the preceding section took the 3 graphs shown in Eq. \Ref{W2SD} we include now all graphs consisting of rings (necklaces) of doubled lines. These are shown in Eq.
 \Ref{Done} in the Appendix in generic form. In the unbroken phase where all
 fields have the same mass and there are no triple vertices, these represent
 all graphs where the summation over the internal indices is to be understood.
 In the broken phase we have to consider both kinds of fields, $\eta$ and $\phi$, and the triple vertices. The number of the correponding graphs becomes larger. The motivation for this choice is that in the unbroken phase  these are all graphs in the next-to-next-to-leading order for large $N$. The leading order is given by the tree approximation, the next-to-leading order by the SDA in the preceding section.

As the expressions in this approximation beyond SDA are quite involved
we are forced to make further approximations. 

The aim of the present section is to find out whether the results
found in the SDA are stable with respect in the next order.  For this
reason we restrict ourselves to the consideration of the SD equations
in the broken phase, Eqs. \Ref{geq1a} and \Ref{geq1b}, which we
rewrite here in the form
\bea\label{sdeq}
M_\eta^2&=&2\tilde{\Gamma}\left(
m^2-\frac{
3\la}{N}\Sigma_\eta^{(0)}-\la\frac{N-1}{N}\Sigma_\phi^{(0)}+2\Gamma
\right) ,
 \nn \\
M_\phi^2&=& \frac{2\la}{N}\left(
\Sigma_\phi^{(0)}-\Sigma_\eta^{(0)}
\right)-gM_\eta^2+f,
\eea
where the notations 
\bea\label{gt}\tilde{
\Gamma}&=& 1-\frac{N}{\la} \frac{\delta
\Gamma}{\delta\beta_\eta(0)}, \\ 
g
&=&\frac{N}{\la(N-1)2
\tilde{\Gamma}}\frac{\delta \Gamma}{\delta\beta_\phi(
0)} ,\nn \\ 
f&=& 2\left(
\frac{\delta D}{\delta\beta_\eta(0)} 
-
\frac{1}{N-1}\frac{\delta D}{\delta\beta_\phi(0)}\right)  \nn
\eea
have been introduced. The functions $\frac{\delta D}{\delta\beta_
\phi(0)}$ and $\Gamma=\frac{\delta D}{\delta\beta_\eta(0)}$ entering here,
 consist of the graphs shown in the Appendix, Eqs. \Ref{Gamma} and \Ref{Dgen}. The corresponding analytic expressions are given by Eqs. \Ref{Gamma1} and \Ref{Df}.  Now we need to take the high-$T$ approximation  of these functions.  For this reason we consider their generic form. Consider, for example, the
fragment
\[ \Tr_q \beta_\phi(q) aA=\Tr_q 
 \beta_\phi(q) 
 \frac{\left(\frac{-3\la}{N}
\Sigma_\eta^{(1)}(q)\right)
^2}{1+\frac{3\la}{N}\Sigma_\eta^{(1)}(q)},
\]
entering Eq. \Ref{gamma2}. Note, that the external momentum is zero as we consider the gap equations. Now we approximate the function $A$, Eq. \Ref{A}, entering here, by a constant, namely its value at zero momentum, $A\to A_{|_{q=0}}$. As $A$ is a dimensionless function which does not change its sign, in the remaining integral the behavior of theintegrand is changed only slightly. Then we take the high-$T$ approximation. The quantities $a$, $b$ and $c$ can be 
calculated in a standard way, see e.g. $\Sigma_1(p)$ in Eq. (57) in
\cite{Bordag:2000tb},
\bea\label{hTa}
a&=& \frac{-3\la T}{8\pi N}        
     \frac{1}{M_\eta} +\dots  \, , \nn\\
b&=& \frac{-3\la T}{8\pi N}
\frac{N+1}{3}\frac{1}{M_\phi} \ +\dots \, ,\nn\\
c&=& \frac{-3\la T}{
8\pi N}\frac{2}{3}  \frac{2}{M_\eta+M_\phi}  +\dots \, ,
\eea
where the dots indicate contributions suppressed by powers of $M/T$.
The remaining expressions correspond to graphs of the basket ball 
type. They have been calculated frequently. In the high-$T$ approximation they read
 \bea
\label{2loop}
\Tr_q \beta_\eta(q) a & =&
\Tr_q \beta_\eta(q) \frac{-3\la}
{N}\Sigma_\eta^{(1)}(q)
= \frac{-3\la T^2}{32\pi^2 N}\ \gamma_1 +\dots  
   \nn\\
\Tr_q \beta_\eta(q) b & =&
\Tr_q \beta_\eta(q) \frac{-3\la}{N}
\frac{N+1}{3}\Sigma_\phi^{(1)}(q)
= \frac{-3\la T^2}{32\pi^2 N}\frac{N+1}
{3} \ \gamma_2 +\dots     \nn\\
\Tr_q \beta_\eta(q) c& =&
\Tr_q \beta_\eta(
q) \frac{-2\la}{N}\Sigma_{\eta\phi}^{(1)}(q)
= \frac{-3\la T^2}{32\pi^2 N}
\frac23  \ \gamma_4 +\dots     \nn\\
\Tr_q \beta_\phi(q) a & =&
\Tr_q \beta_
\phi(q) \frac{-3\la}{N}\Sigma_\phi^{(1)}(q)
= \frac{-3\la T^2}{32\pi^2 N}\
 \gamma_4 +\dots     \nn\\
\Tr_q \beta_\phi(q) b & =&
\Tr_q \beta_\phi(q)
 \frac{-3\la}{N}\frac{N+1}{3}\Sigma_\phi^{(1)}(q)
= \frac{-3\la T^2}{32\pi^2
 N}\frac{N+1}{3} \ \gamma_3 +\dots     \nn\\
\Tr_q \beta_\phi(q) c& =&
\Tr_q \beta_\phi(q) \frac{-2\la}{N}\Sigma_{\eta\phi}^{(1)}(q)
=
 \frac{-3\la T^2}{32\pi^2 N}\frac23  \ \gamma_2 +\dots     
\eea
with
\bea\label{logs}
\gamma_1&=& 1-2\ln \frac{3M_\eta}{\mu} \ , 
\nn\\
\gamma_2&=& 1-2\ln \frac{M_\eta+2M_\phi}{\mu}\ ,  \nn\\
\gamma
_3&=& 1-2\ln \frac{3M_\phi}{\mu}\ ,  \nn\\
\gamma_4&=& 1-2\ln \frac
{2M_\eta+M_\phi}{\mu} \ , 
\eea
where $\mu$ is a normalization constant. Again, by the dots higher
orders in $M/T$ are indicated.

In this way, we have analytic expressions for the functions entering
the gap equations \Ref{sdeq} which can be written as 
\bea\label{Gammaapp}
\Gamma&=&\frac{-3\la^2T^2}{32\pi^2N^2}
\Big\{
        \gamma_1(1+3A)+(
N+1)\gamma_2(\ep+B)\nn   \\ && \left. ~~~~~~~~~
+3\frac{\gamma_1}{a}
\frac{(2+a+B)AB}{1-AB}
+\frac23(N-1)\gamma_2\frac{c}{1-c}
\right\},
\eea
\bea\label{Gaeapp} \frac{\delta \Gamma}{\delta
\beta_\eta(
0)}&=&\frac{-\la}{N}\left\{ \left(1+2a\pa_a\right)\left[
a(1+3A)+3
\frac{(2+a+B)AB}{1-AB}\right] \right. \nn \\ &&
~~~~~~~~\left. +2
\frac{N-1}{N+1}b\pa_c\frac{c^2}{1-c}\right\} \eea
and
\bea\label
{Dfapp} \frac{\delta D}{\delta
\beta_\phi(0)}&=&\frac{-3\la^2T^2}{32
\pi^2N^2} \left\{
\frac{(N+1)^2}{3}\gamma_3\left[\frac13+\frac{b}{1-b}
+2\frac{N-2}{N+1}\left(\frac13+\frac{2}{N+1-2b}\right)\right]
\right.
 \nn \\ && ~~~~~~~~
+(N-1)\left[\frac{1+A}{3}\gamma_4+\frac{\gamma_4}
{3\ep
a}\frac{\left(2+\ep A+\frac{1}{\ep}B\right)AB}{1-AB} \right] \nn
 \\ &&
~~~~~~~~ \left.  +(N-1)\frac{2\gamma_2}{3}\frac{c}{1-c} \right\}
 \eea
with $a$, $b$ and $c$ given by Eqs. \Ref{hTa}, for the remaining
notations see the Appendix.

The functions $\Sigma^{(0)}$ entering the gap equations have been given for high $T$ in the preceding section by Eqs. \Ref{ge2}. Here, in order to be consistent with \Ref{2loop}, we include one more term into the expansion, i.e., we take
\be\label{} \Sigma^{(0)}=\frac{T^2}{12}-\frac{MT}{4\pi}+
\frac{M^2
L_n+2m^2}{16\pi^2}+\dots \ee
with the notation 
\be\label{Ln}L_n=\ln\frac{(4\pi T)^2}{2m^2}-2\gamma,
\ee
where $\gamma$ is Euler's constant gamma. 

Inserting these approximations we rewrite the gap equations \Ref{sdeq} in the form
\bea\label{sdeq2} M_\eta^2&
=& 2h\left( m^2\delta_1-\frac{\la
T^2}{4N}\delta_\gamma+D_\phi +M_\eta
\frac{3\la T}{4\pi N}\right), \nn
\\ M_\phi^2&=& \frac{\la T}{2\pi
N}
\frac{-M_\phi^2}{M-\eta+M_\phi}+\frac{\la L_n}{8\pi^2 N} _g \
M_\eta^2
+f \eea
with
\bea \label{adds} D_\phi&=& \frac{\la(N-1)T}{4\pi
N}M_
\phi-\frac{\la(N-1)L_n}{16\pi^2 N}M_\phi^2, \nn \\ h&=&
\frac{\tilde{
\Gamma}}{1+2 \tilde{\Gamma}\frac{3\la L_n}{16\pi^2 N}},
\nn \\ \delta_
1&=& 1-\frac{3\la}{8\pi^2N}\frac{N+2}{3}, \nn \\
\delta_\gamma&=& 
\frac{N+2}{N}-\frac{3\la }{4\pi N}
\left(\frac{N+2}{N}-2\ln
\frac{3M_
\eta}{\mu}-s\ln\frac{M_\eta+2M_\phi}{\mu}\right).
\eea
The quantity $D_\phi$, entering the first gap equation, depends on
$m_\phi$ which must be found from the second equation as a function of
$m_\eta$.

Now, we investigate qualitatively these equations and plot both sides in
one plot. The lhs. is simply part of a parabola. The rhs. have a more
complicated behavior. For large $M_\eta$ resp. $M_\phi$ they remain smaller than the lhs because of additional factors $\la$ in front. For small $M_\eta$ in the rhs. of the first equation we note
\bea
\label{smM} h&=& \frac{8\pi^2 N}{3\la L_n}+O(M_\eta), \nn \\
\delta_
\gamma&=&\frac{3\la}{\pi N} \ln\frac{1}{M_\eta}+O(M_\eta),
\eea%
so that it takes negative values for small $M_\eta$. 

It is interesting to compare with the expansion of these quantities for small $\la$. This corresponds to take the first graph out of the infinite sequence
 \Ref{Done}. In the SD equation this means to take the graph 
\epsfxsize=1.5cm
\raisebox{-0.2cm}{ \epsffile{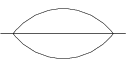}} in addition
 to $\Sigma^{(0)}$, Eq. \Ref{dfs}, in the self energy part. As an
approximation, this is in between SDA and BSAD and does not include
the effect from summing up the infinite chain of doubled lines.  We
call this perturbative approximation beyond the super daisy approximation (pBSDA). So, in pBSDA it holds
\bea\label{pbsad}
 h&=&1+\dots   \\ \delta_\gamma&=&
\frac{N+2}{3}\left(1-\frac{3\la}{4\pi
 N}\right)+\frac{3\la}{2\pi
N}\left(\ln\frac{3M_\eta}{\mu}+\frac{N-1}{3}
\ln
\frac{M_\eta+2M_\phi}{\mu}\right)
+\dots \ . \nn \eea
Note, that $\delta_\gamma$ takes for sufficiently small $M_\eta$ negative values in opposite to \Ref{smM}.

\begin{figure}[h]\unitlength=1cm 
\begin{picture}(5,6.0)
\put(0,0)
{ 
\epsffile{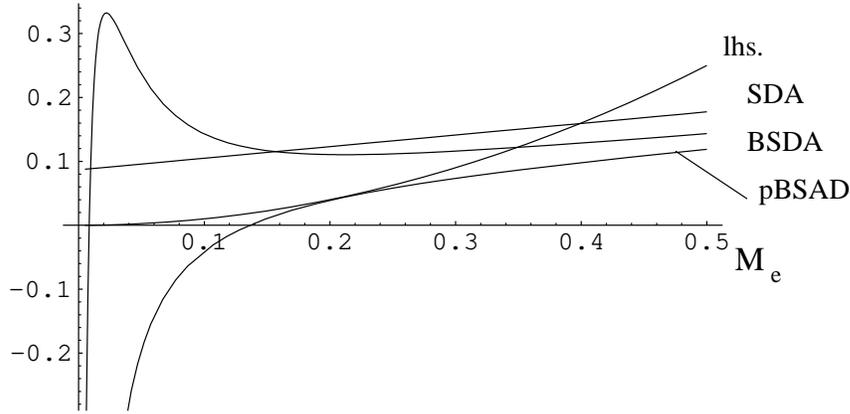}}
\end{picture}
\caption{The lhs. and the rhs. of the
 first gap equation, \Ref{sdeq2}, for different approximations as function
of the mass $M_\eta$ for $\la=0.1$, $T=7.7$, $M_\phi=0.2$. }

\label{figrhs1}\end{figure}

The rhs. of the first equation is shown in Figure \ref{figrhs1} in
both approximations for a temperature near the phase transition.  The
curve corresponding to the pBSDA starts from some finite value at $M_
\eta=0$. For small temperature ($T<T_-$ where $T_-$ has to be taken in the
 given approximation) it starts from positive values and has one intersection  with the lhs. corresponding to one extremum only of the free energy. In opposite, for $T>T_-$ it starts from negative values and has two intersections indicating the existence of a maximum.  For
larger $T$,
 the curve lowers until its two intersections with the
lhs. merge at the
 corresponding $T_+$. This case is just shown in the figure. 
 This is
 qualitatively the same
behavior as in the SDA case. Now consider the
 BSDA case. Here, due to
the logarithmic behavior in Eq. \Ref{smM}, the
 curve starts for any
temperature from negative values and we have a second
 extremum (which
must correspond to a maximum) for any temperature $T<T_+$.
 So, formally, we
observe $T_-=0$. However, these negative values to be
 realized the
logarithm must be large, $\ln M_\eta \sim -\frac{\pi^2 N}{
3\la}$. This
is the only qualitative change coming in from the BSDA
 resummation and
it is of extreme smallness. In general, it is a quite
 remarkable fact
that the resummation results in a denominator (combinations
 $1-a$,
$1-b$, $1-c$ and $1-AB$) which does not have a zero in the
 region we
are interested in.
Comparing the curves in Figure \ref{figrhs1},
 we see that the
quantitative difference between the different approximation
 is quite
large.

\begin{figure}[h]\unitlength=1cm 
\begin{picture}(5,6)
\put(0,0){ 

\epsffile{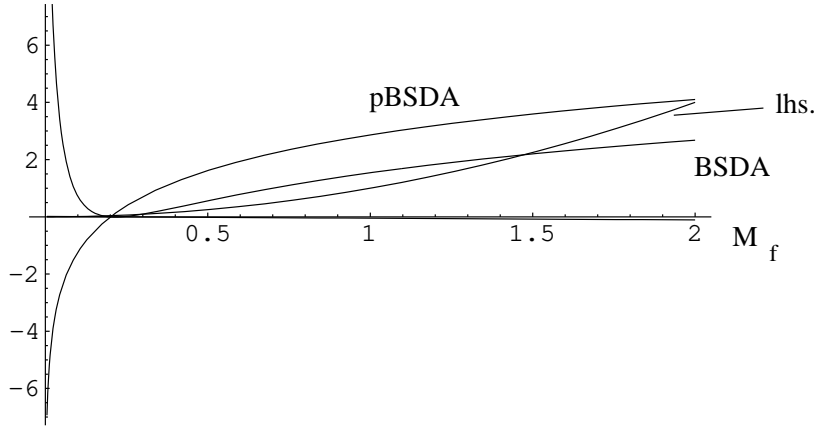}}
\end{picture}
\caption{The lhs. and the rhs. of the
 second gap equation, \Ref{sdeq2}, 
for different approximations as
 function
of the mass $M_\phi$ for $\la=0.1$, $T=7.7$, $M_\eta=0.2$. }

\label{figrhs2}\end{figure}

Let us now consider the second equation. It is shown in
 Figure
\ref{figrhs2}.  As one can see, in any case in both approximations 
 it has a solution which is similar to the simple solution of the
 second
equation in the SDA case, Eq. \Ref{Mf}. However, for some value
s of
the parameters, there may be three solutions as seen from the
figure.
 This is the substructure of the phase transition which deserves
further
 investigation.

\begin{figure}[h]\unitlength=1cm\label{figBSD}
\begin{picture}(5,13)

\put(0,8.5){ \epsffile{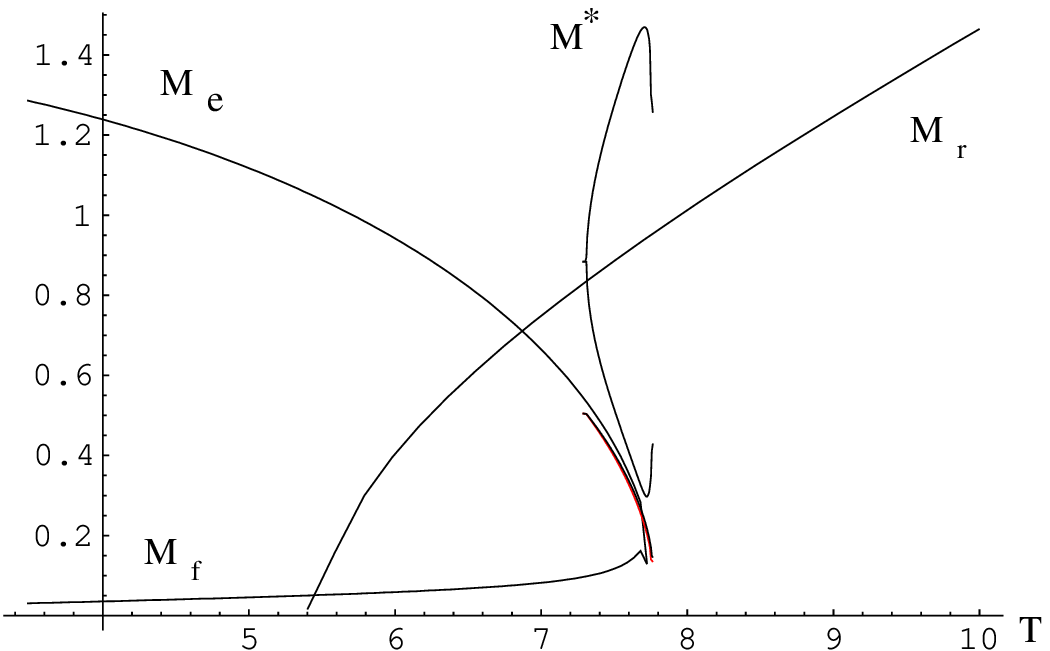}}
\put(0,-1){ \epsffile{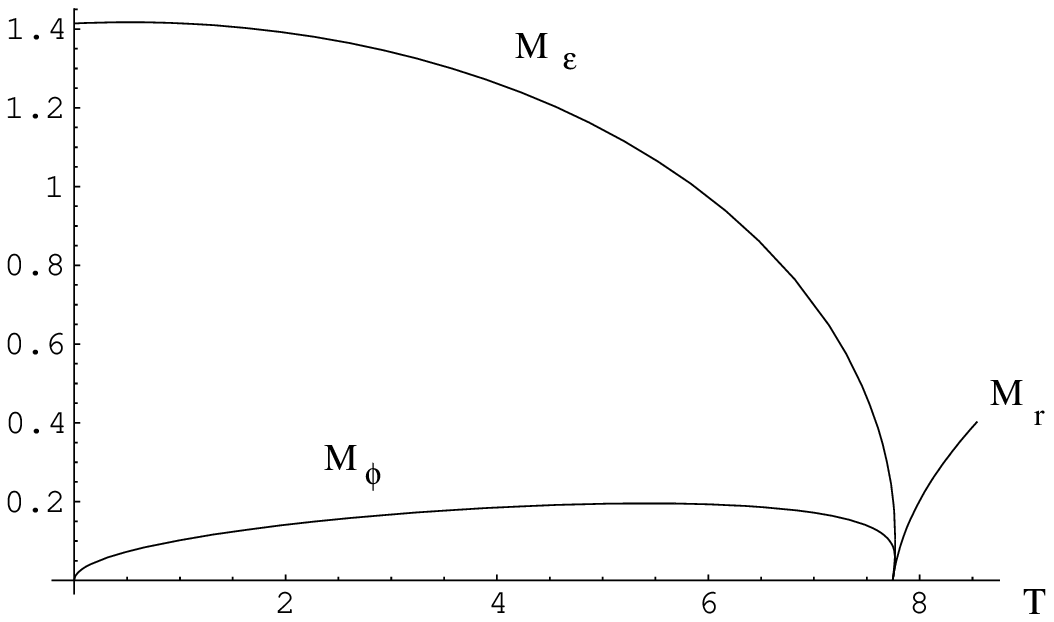}}
\end{picture}
\caption{The upper plot shows the numerical solution of
 the gap
equations in the BSDA case as function of the temperature $T$
 for
$\la=0.1$. Here $M_\eta$, $M_\phi$ are the masses in the broken phase
in the minimum of the free energy (the corresponding masses in the
maximum
 are not shown), $M_r$ is the mass in the restored phase and
$M^*$ is the
 mass for the additional solutions.  The lower plot shows
for comparizon the
 corresponding solution in the SDA case. }
\end{figure}

Having clarified the general behavior of the solutions of
 the gap
equations, one can start a numerical investigation. Results are
 shown
in backing the qualitative similarity to the SDA.  A numerical
solution in the BSDA case is shown in Figure \ref{figBSD}. It shows
qualitatively the same behavior as in the SDA case which is shown in
the lower plot in Figure \ref{figBSD} for the same value of $\la=0.1$. We note that the solution in the restored phase starts in the BSAD case at considerably smaller $T$. In fact it starts from $T=0$ but there it is exponentially small.

\section{Conclusions}

We investigated the phase transition in the $O(N)$-model by
perturbative methods. As technical tools we used the second Legendre transform and the  method of considering the gap equations in the extrema of the free energy. The latter simplifies the expressions considerably and allows for more
 explicit results.
 
 We considered two stages of approximation. We started with the super daisy
 approximation. Here, for small coupling $\la$, which is by means of Eq.
 \Ref{T-sd}, equivalent to high temperature, we obtained clearly a first order
 phase transition. This generalizes the corresponding results obtained in the
 one component case found in \cite{Bordag:2000tb}. The masses of the Higgs and
 Goldstone particles in the broken phase and the mass of the field in the
 restored phase are shown in Figure \ref{figure1}. The correponding free
 energy is shown in Figure \ref{figure2}. It should be noticed that the first
 order character of the phase transition is extremely weak. As can be seen
 from Figure \ref{figure2}, the change in the free energy between $T_c$ and
 $T_-$ is numerically small as compared to the mass scale which is set to unit
 in that figure. Also, the difference between the two spinodal temperatures,
 Eq. \Ref{T+la}, is of order $\la$. So, to leading order in $\la$ the
 transition is of second order and the first order character is a
 next-to-leading effect in $\la$.  For large $N$, from Eq. \Ref{tlargeN} it
 follows that the transition becomes second order.
 
 In order to check the reliability of the results obtained in SDA we
 calculated in Section 4, beyond the SDA, the next class of graphs (BSAD). In
 the unbroken phase these are all graphs in the next-to-next-to-leading order
 in $1/N$. They are given by Eq. \Ref{Done} and their dependence on $N$ is
 taken into account completely (not only for large $N$).  These graphs can be
 characterised as chains of doubled lines.  The summation of the infinite set
 of these graphs is performed for the quantities entering the gap equations.
 Then the resulting quantities have been calculated approximatively for high
 temperature where in essence besides the leading contribution only the
 logarithmic terms, Eq. \Ref{logs}, have been taken into account. As the
 remaining expressions are quite complicated, the gap equations have been
 solved numerically. An example for such a solution is shown in Figure
 \ref{figBSD}. The general outcome is that the results obtained in SDA remain
 valid qualitatively. They are altered in details. For instance, there exists
 a maximum of the free energy for any $ T<T_+$ so that formally we have
 $T_{-}=0$. However, for small $T$ this maximum is exponentially small (it is
 due to the logarithmic terms). Also, there appear additional solutions ($M^*$
 in Figure \ref{figBSD}).  In this way we confirmed in the $N$-component case
 the results found in \cite{Bordag:2000tb} for the one component case stating
 that the results in SAD are qualitatively stable with respect to higher
 loops.

 With these results obtained we are left with the difficult problem that all nonperturbative approaches (lattice calculations, flow equations, etc.) predict a second order character for the transition. In the absense of a small expansion
 parameter near the phase transition it was resonable to believe that the origin of the discrepancy between the perturbative and non perturbative results is due to this fact. If this is the case, it is important to find out the type of diagrams responsible for the change of the type of the phase transition. 
 
 Now, we are going to comment some calculations where the second order phase
 transition has been determined. First we discuss the observation in
 Ref.\cite{Ogure:1998xu} that the inclusion of a two-loop diagram beyond the
 super daisy resummation results in a second order phase transition. The
 diagram mentioned is the first one in a series depicted in Eq. \Ref{Gamma} in
 the Appendix. Its exclusion does not change significantly the structure of
 the gap equations and the solutions obtained if, of course, all other
 diagrams are taken into consideration. Hence, we conclude that this diagram
 does not change the type of the phase transition. In
 Ref.\cite{Elmfors:1992yn} the second order phase transition was observed when
 the renormalization group method at finite temperature has been applied.
 Results obtained in this way are difficult to compare with that found in the
 case of the standard renormalization at zero temperature adopted in the
 present paper. The point is that the renormalization at finite temperature
 replaces some resummations of a series of diagrams which remains unspecified
 basically.
 
 In Refs.\cite{Ogure:1998je,Ogure:1998xu} new non perturbative method of
 calculation of the effective potential at finite temperature - an auxillary
 field method - has been developed and a second order phase transition was
 observed for both the one- and $N$-component models. We left the detailed
 analysis of this method for other publication and here just mention that it
 seems to us not a self-consisten one because it delivers an imaginary part to
 the effective poterntial in its minima. This important point is crucial for
 any calculation scheme as a whole. Really, the minima of the effective
 potential describe physical vacua of a system. An imaginary part is signaling
 either the false vacuum or the inconsistency of the calculation procedure
 used. That is well known starting from the work by Dolan and Jackiw
 \cite{dola74-9-3320} who noted the necessity of resummations in order to have
 a real effective potential at finite temperature. The key point of the
 procedure in Refs.\cite{Ogure:1998je,Ogure:1998xu} is to relate the total
 polarization operator and the one-loop effective potential. But it is
 difficult to achive on some resonable grounds because these entities are
 different parts of the effective action in the derivative expansion. Hence,
 it is difficult to estimate the correctness of the method, its stability with
 respect to some small deviations, etc. In our method of calculations the
 stability of the vacuum is authomatically satisfied when the super daisy
 diagrams are resummed in the extrema of the free energy.

As we have seen, the SDA possesses a lot of attractive features. In particular, the results obtained in this way are qualitatively stable with respect to other resummations. So, it is resonable to investigate in more detail its properties as well.


\section*{Appendix}
In this appendix we calculate and sum up the bubble chain graphs
appearing in the gap equations in sections 2 and 4. The corresponding well known Feyman rules are collected in Table 2.
\begin{table}
\begin{tabular}{rcc}
graphic element   & corresponding field  & propagator\\[3pt]\hline \\[0.1cm]
\epsfxsize=4cm\epsfysize=0.1cm\epsffile{lineeta.ps}  &$ \eta$     & $\beta_\eta=(p^2+M_\eta^2)^{-1}$\\[6pt]    
\epsfxsize=4cm\epsfysize=0.1cm\epsffile{linephi.ps}                 &  $ \phi_a$     & $\beta_\phi=(p^2+M_\phi^2)^{-1}$\\[12pt]
\end{tabular}
\begin{tabular}{ccc}
graphic element   & vertex factor & notation \\[4pt] \hline \\[0.1cm]
\epsfxsize=4cm\epsfysize=4cm\raisebox{-3.3cm}{\epsffile{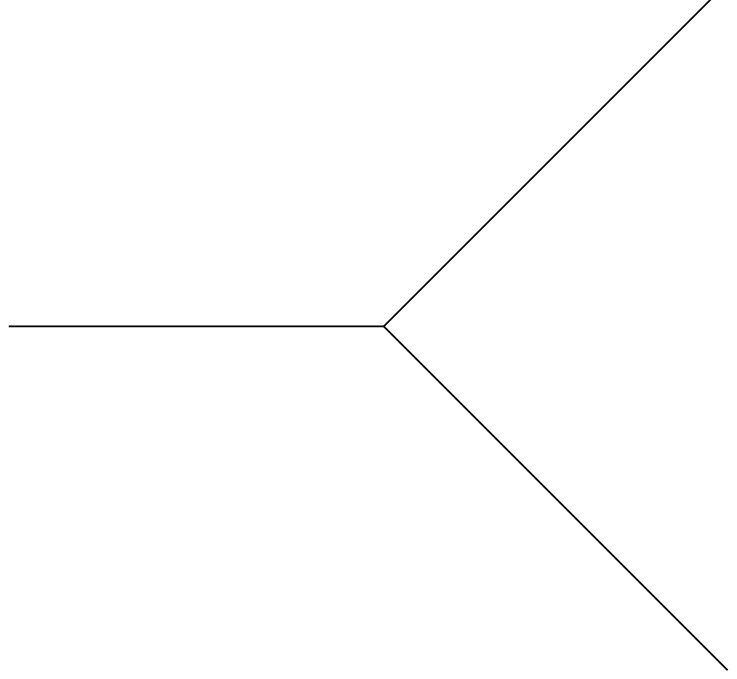}}  &$-\frac{6\la v}{N} $           & ($\eta^3$)      \\[-1.9cm]
\epsfxsize=4cm\epsfysize=4cm\raisebox{-3.3cm}{\epsffile{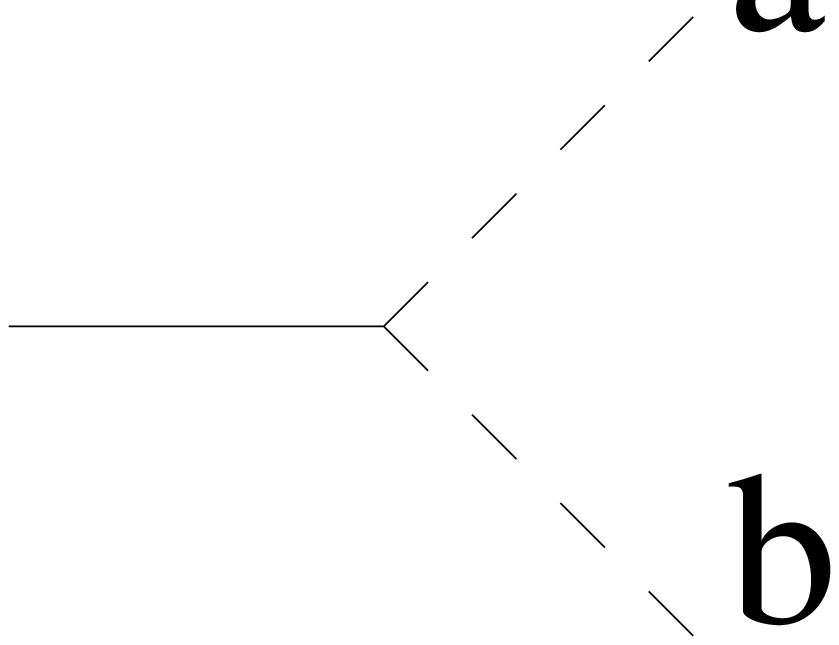}}  &$-\frac{6\la v}{N}\delta_{ab} $ & ($\eta\phi^2$)  \\[-1.9cm]
\epsfxsize=4cm\epsfysize=4cm\raisebox{-3.3cm}{\epsffile{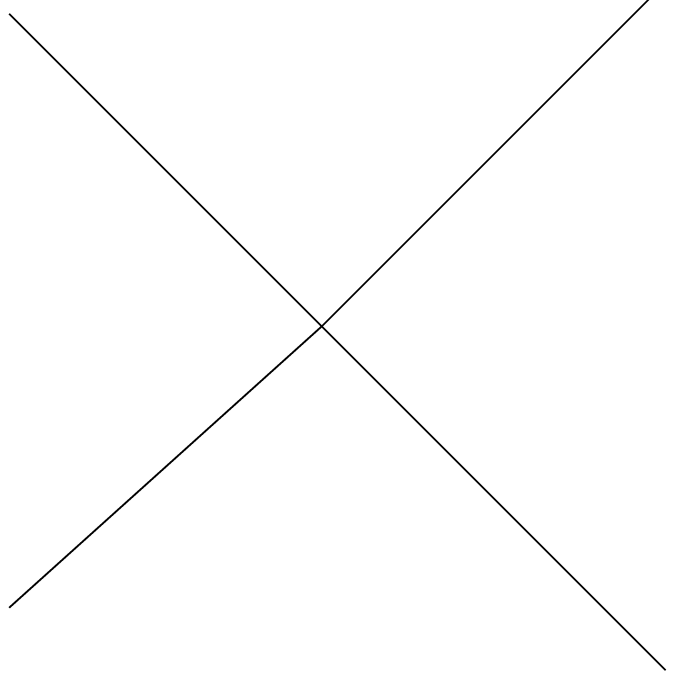}}  &$-\frac{2\la }{N} $          & ($\eta^4$)      \\[-1.9cm]
\epsfxsize=4cm\epsfysize=4cm\raisebox{-3.3cm}{\epsffile{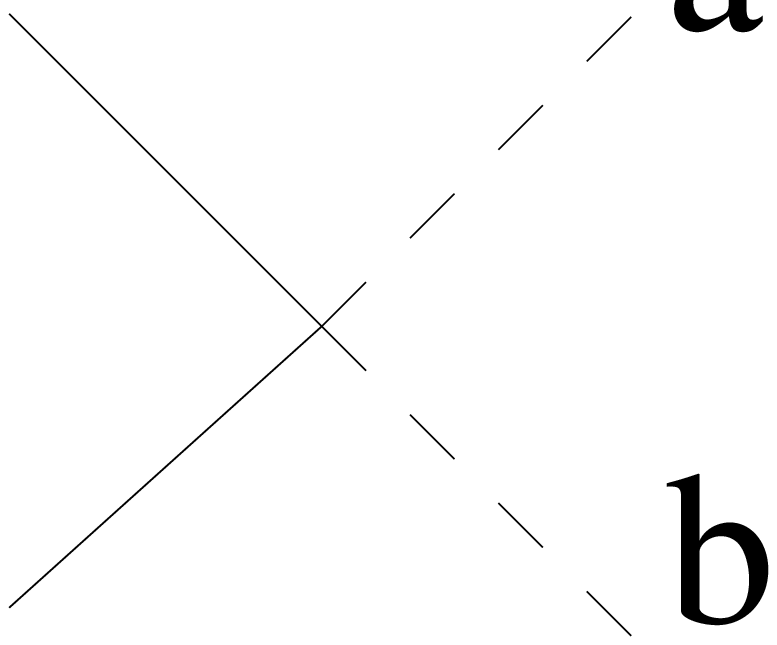}}  &$-\frac{2\la }{N}\delta_{ab}$  & ($\eta^2\phi^2$)\\[-1.9cm]
\epsfxsize=4cm\epsfysize=4cm\raisebox{-3.3cm}{\epsffile{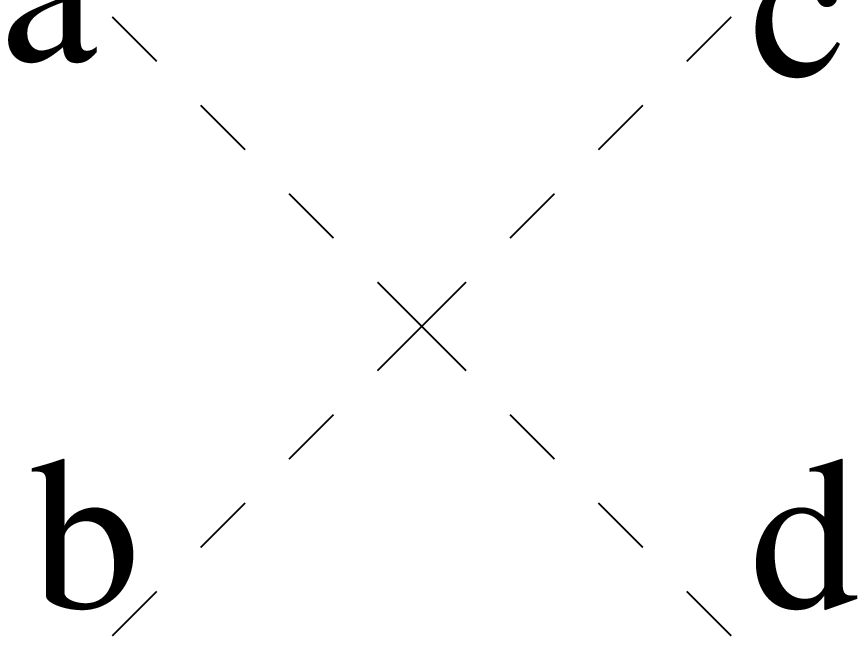}}  &$-\frac{6\la }{N} V_{abcd}$    & ($\phi^4)$ \\[-1.9cm]
 \end{tabular}
\caption{The Feyman rules for the $O(N)$-model}
\end{table}
The vertex $(\phi^4)$ contains the symmetric tensor
\be\label{V}V_{abcd}=\frac13\left(
\delta_{ab}\delta_{cd}+
\delta_{ac}\delta_{bd}+
\delta_{ad}\delta_{bc}   \right).
\ee

The functional $\Gamma$ introduced in \Ref{W2app0} consists of all
graphs with 2 triple vertices, $(\eta^3$) or ($\eta\phi^2$), i.e., of
all graphs proportional to $v^2$ and having at least 2 vertices. In the
approximation made in Section 5 the graphical representation
is
\vspace{0.3cm}
\bea\label{Gamma}
v^2 \Gamma&=&
\frac{1}{12} \ \epsfxsize=2.6cm\epsfysize=3cm\raisebox{-2.4cm}{\epsffile{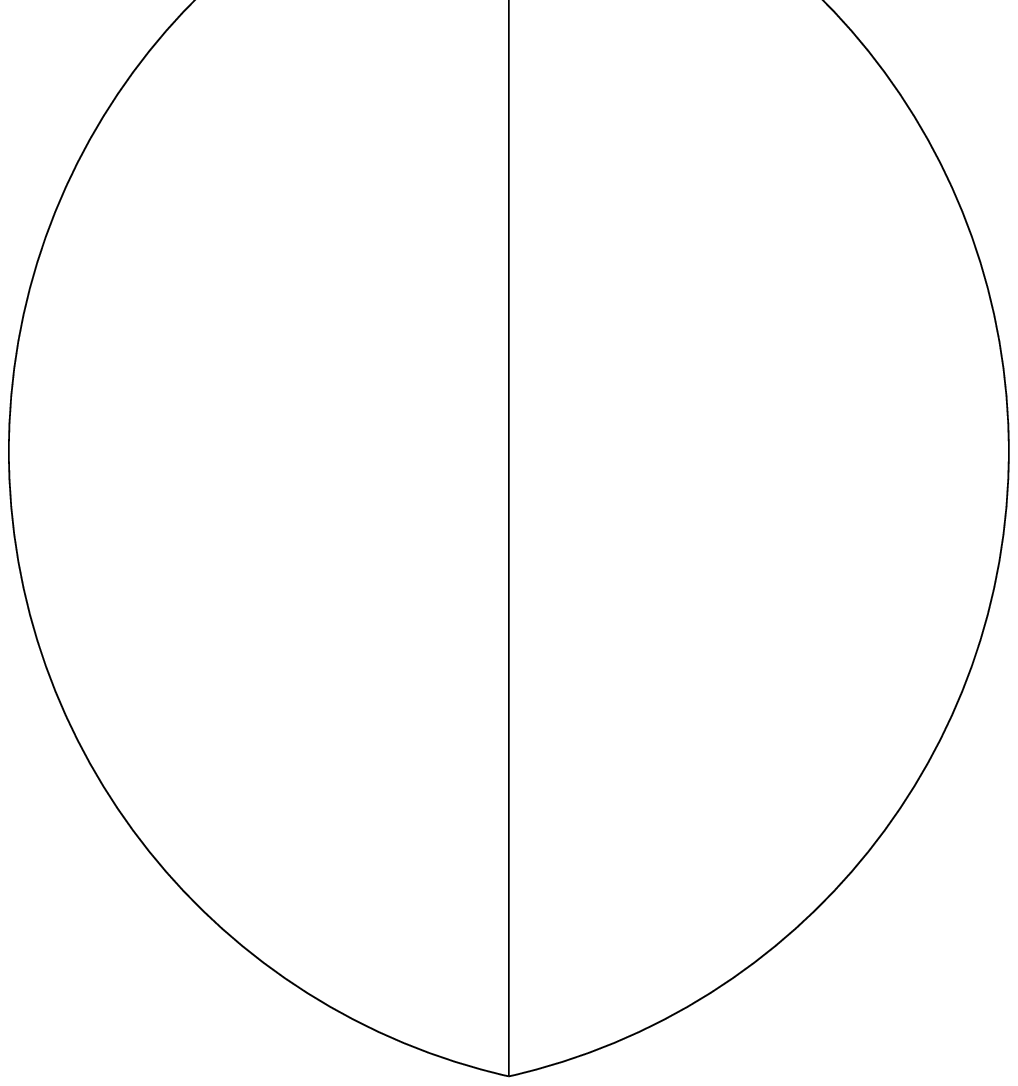}} \hspace{-0.8cm}
+\frac{1}{8} \ \epsfxsize=2.6cm\epsfysize=3cm\raisebox{-2.4cm}{\epsffile{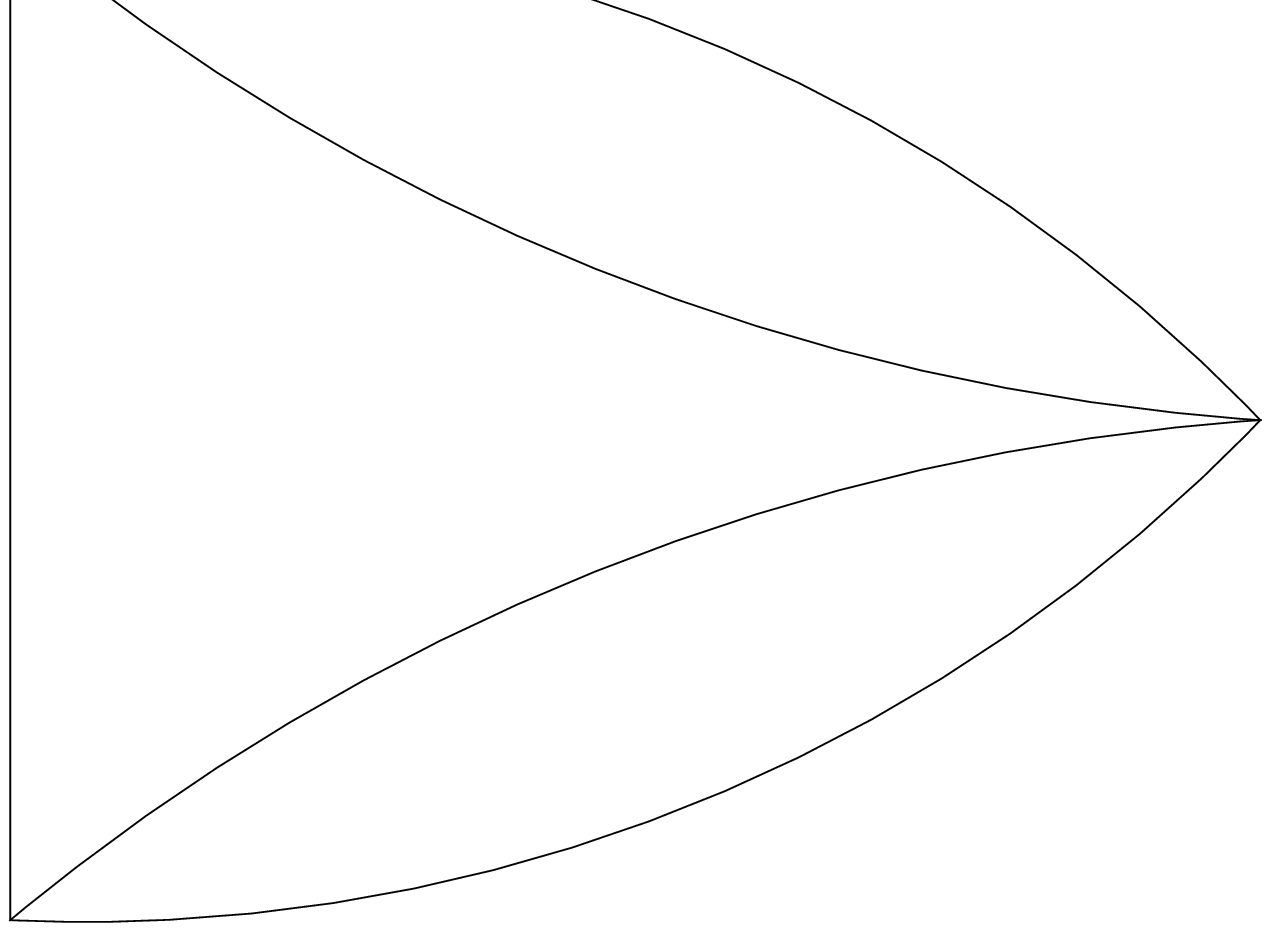}}  \hspace{-0.8cm}
+\frac{1}{16} \ \epsfxsize=2.6cm\epsfysize=3cm\raisebox{-2.4cm}{\epsffile{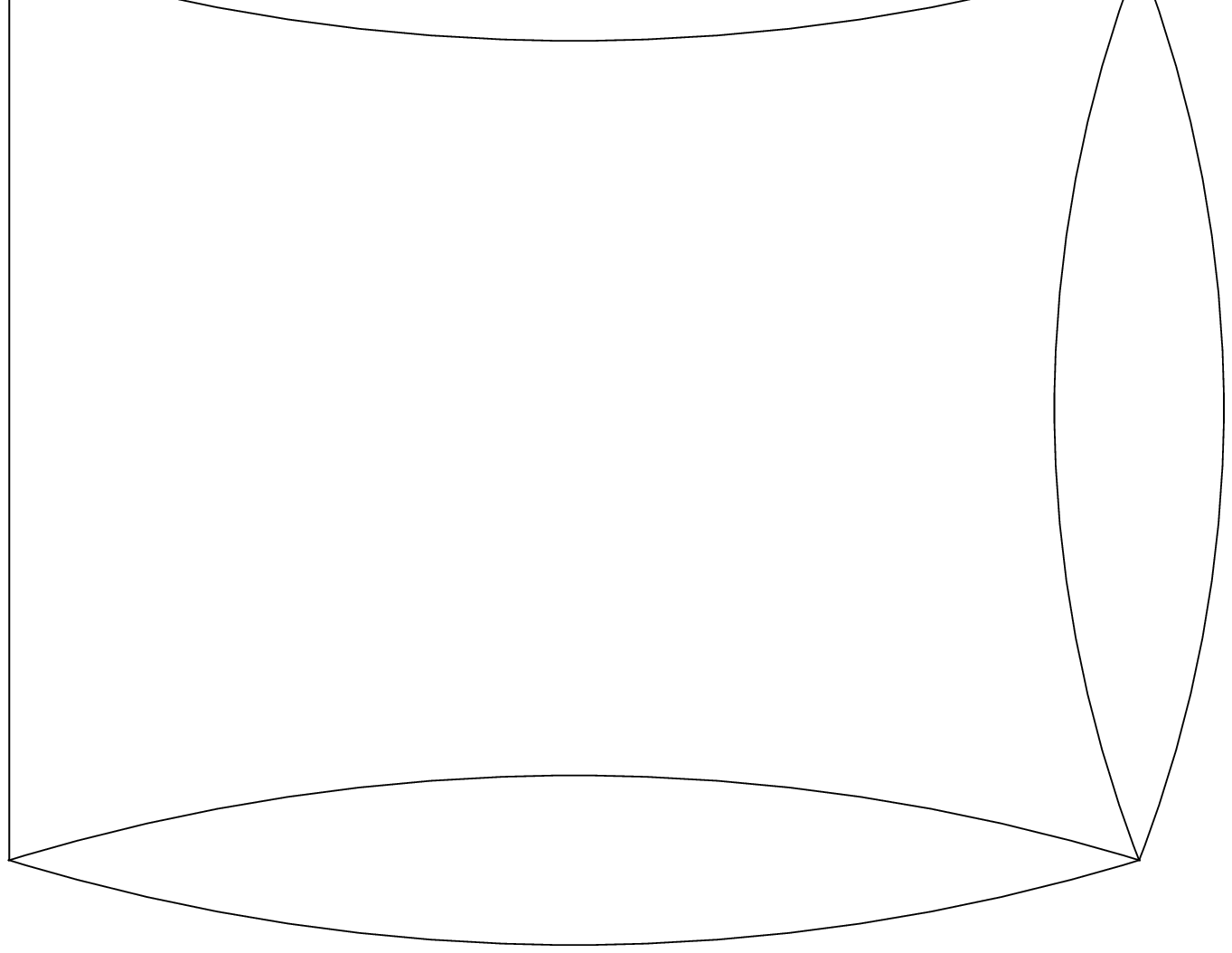}}  \hspace{-0.8cm}
+\dots\nn \\[-0.9cm]&&
+\frac{1}{4} \ \epsfxsize=2.6cm\epsfysize=3cm\raisebox{-2.4cm}{\epsffile{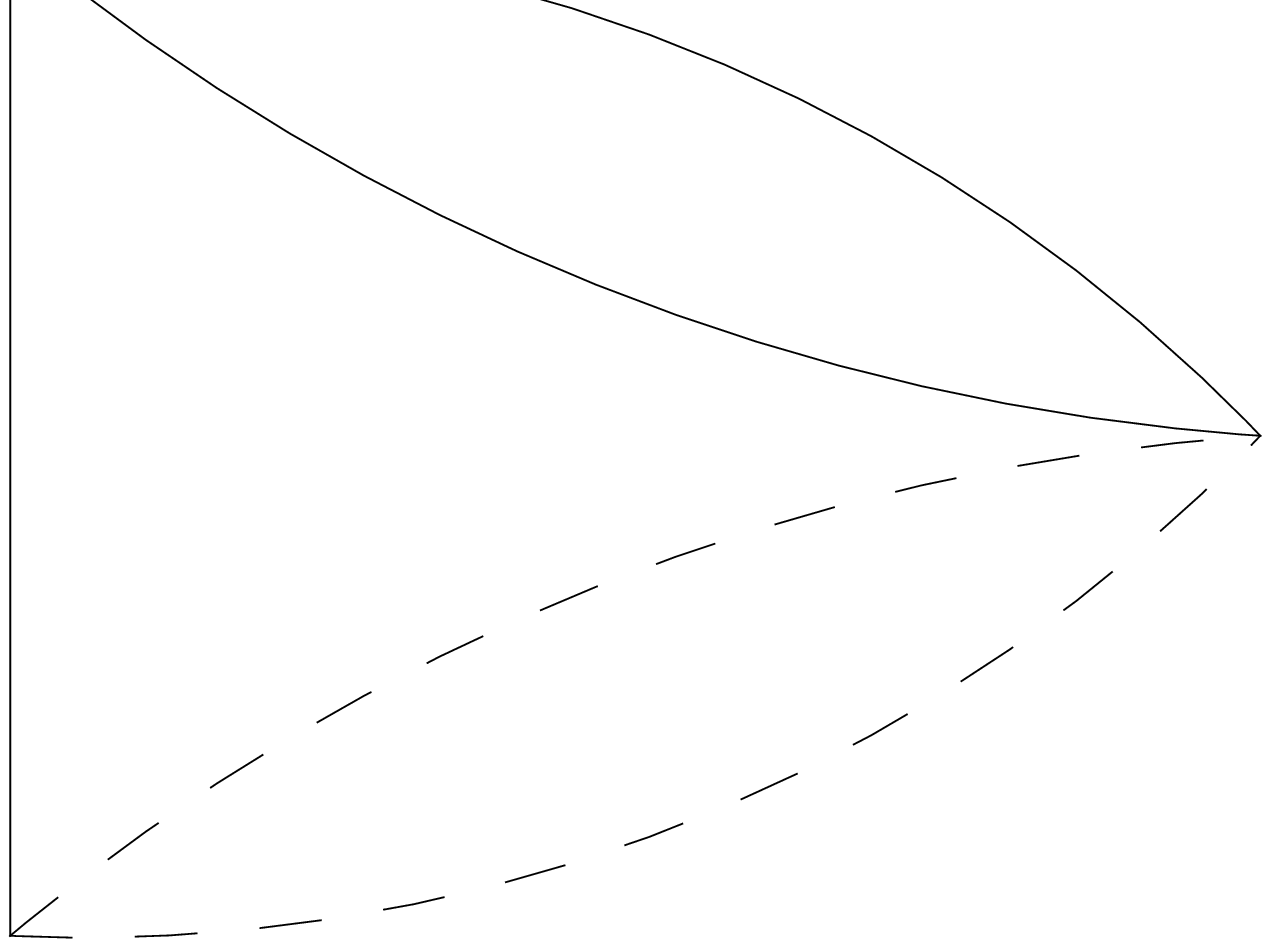}}  \hspace{-0.8cm}
+\frac{1}{8} \ \epsfxsize=2.6cm\epsfysize=3cm\raisebox{-2.4cm}{\epsffile{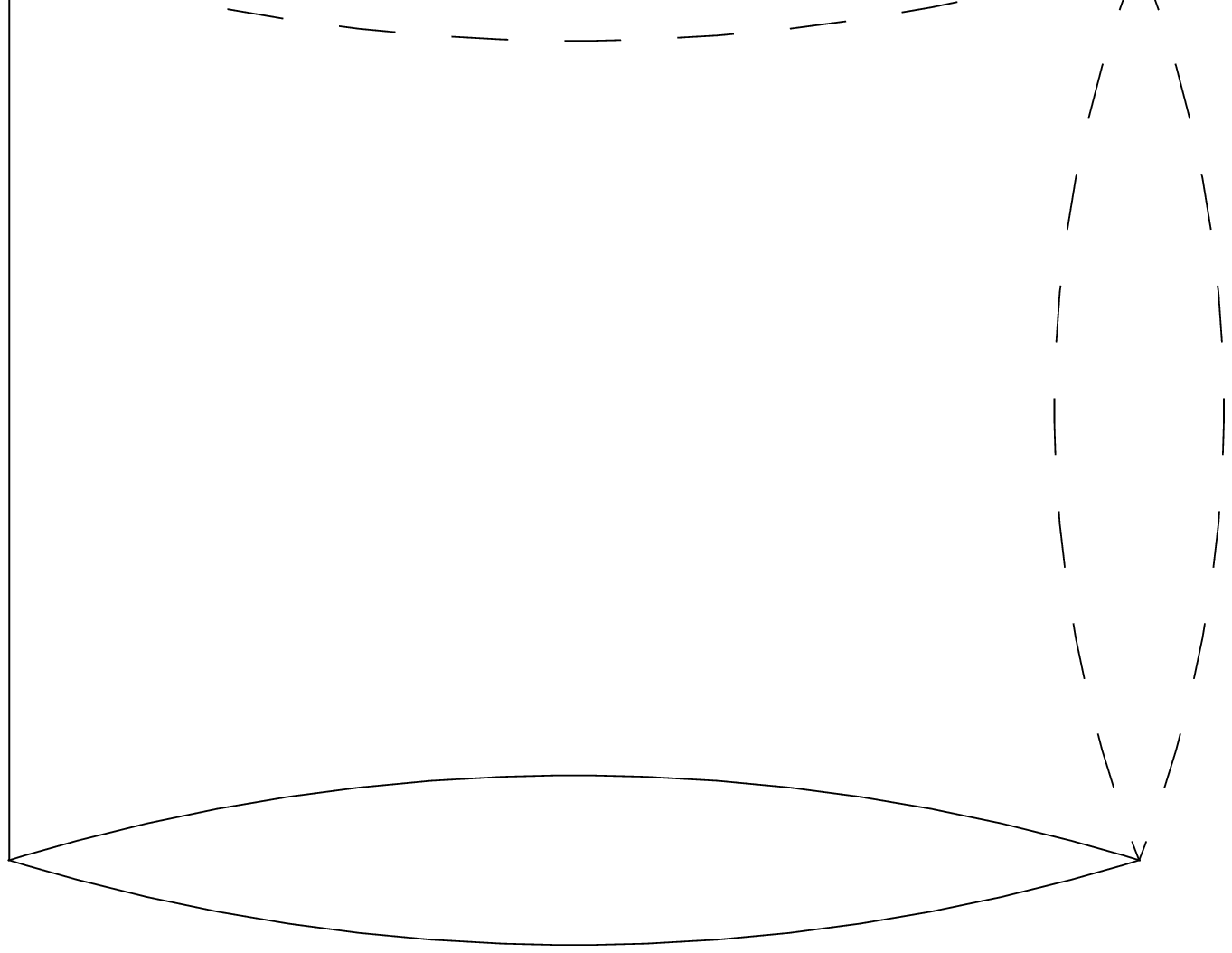}}  \hspace{-0.8cm}
+\frac{1}{8} \ \epsfxsize=2.6cm\epsfysize=3cm\raisebox{-2.4cm}{\epsffile{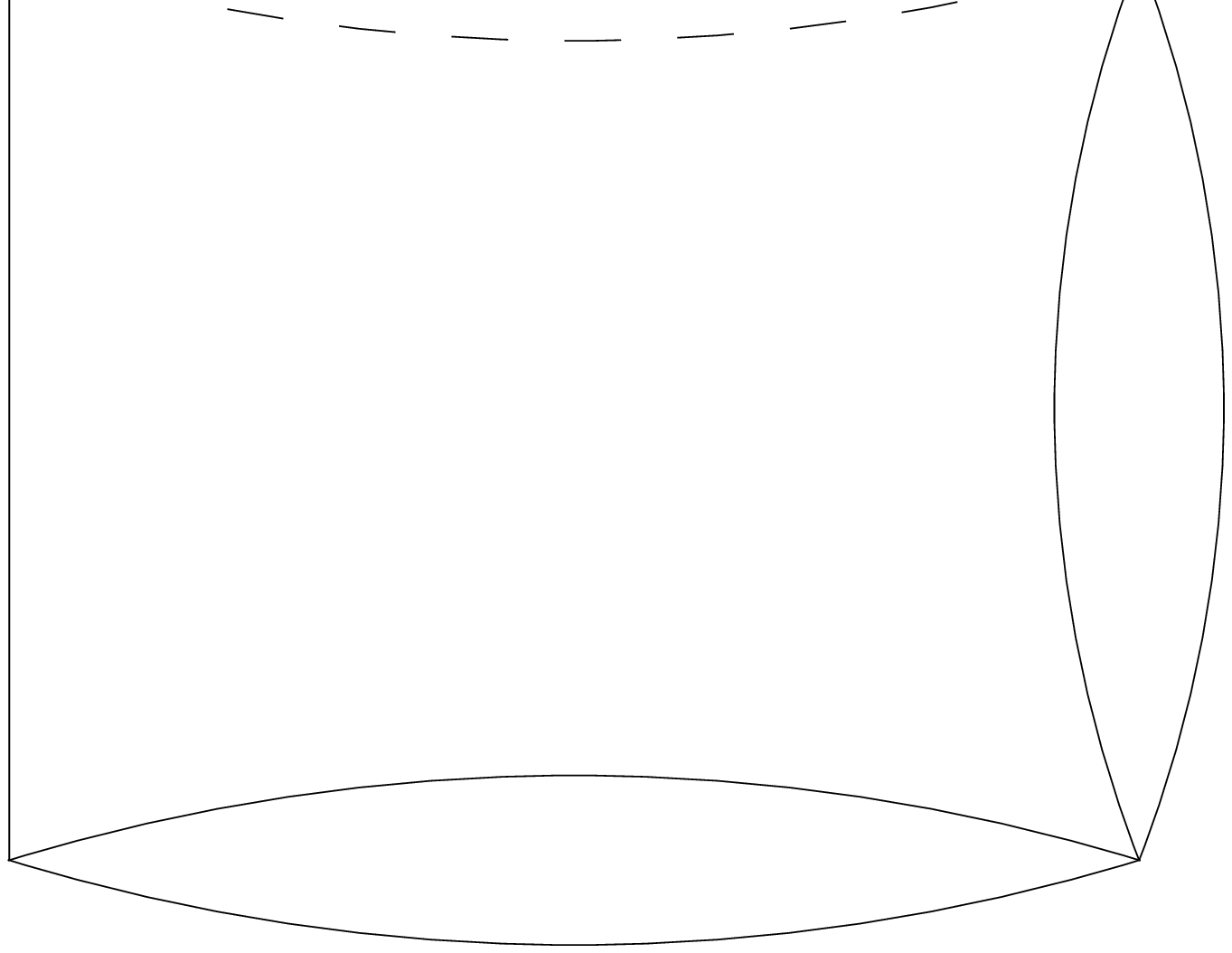}}  \hspace{-0.8cm}
+\dots\nn  \\[-0.9cm]&&
+\frac{1}{4} \ \epsfxsize=2.6cm\epsfysize=3cm\raisebox{-2.4cm}{\epsffile{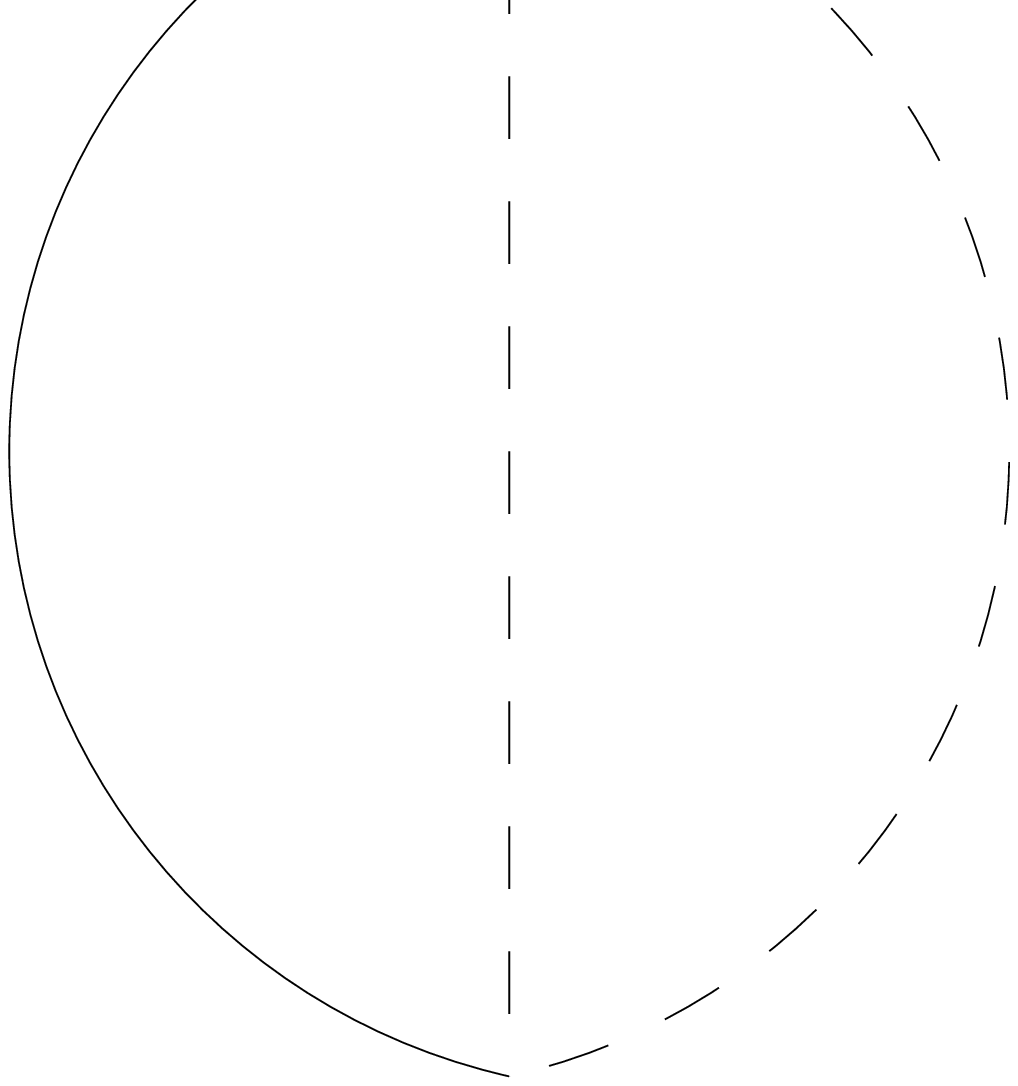}}  \hspace{-0.8cm}
+\frac{1}{8} \ \epsfxsize=2.6cm\epsfysize=3cm\raisebox{-2.4cm}{\epsffile{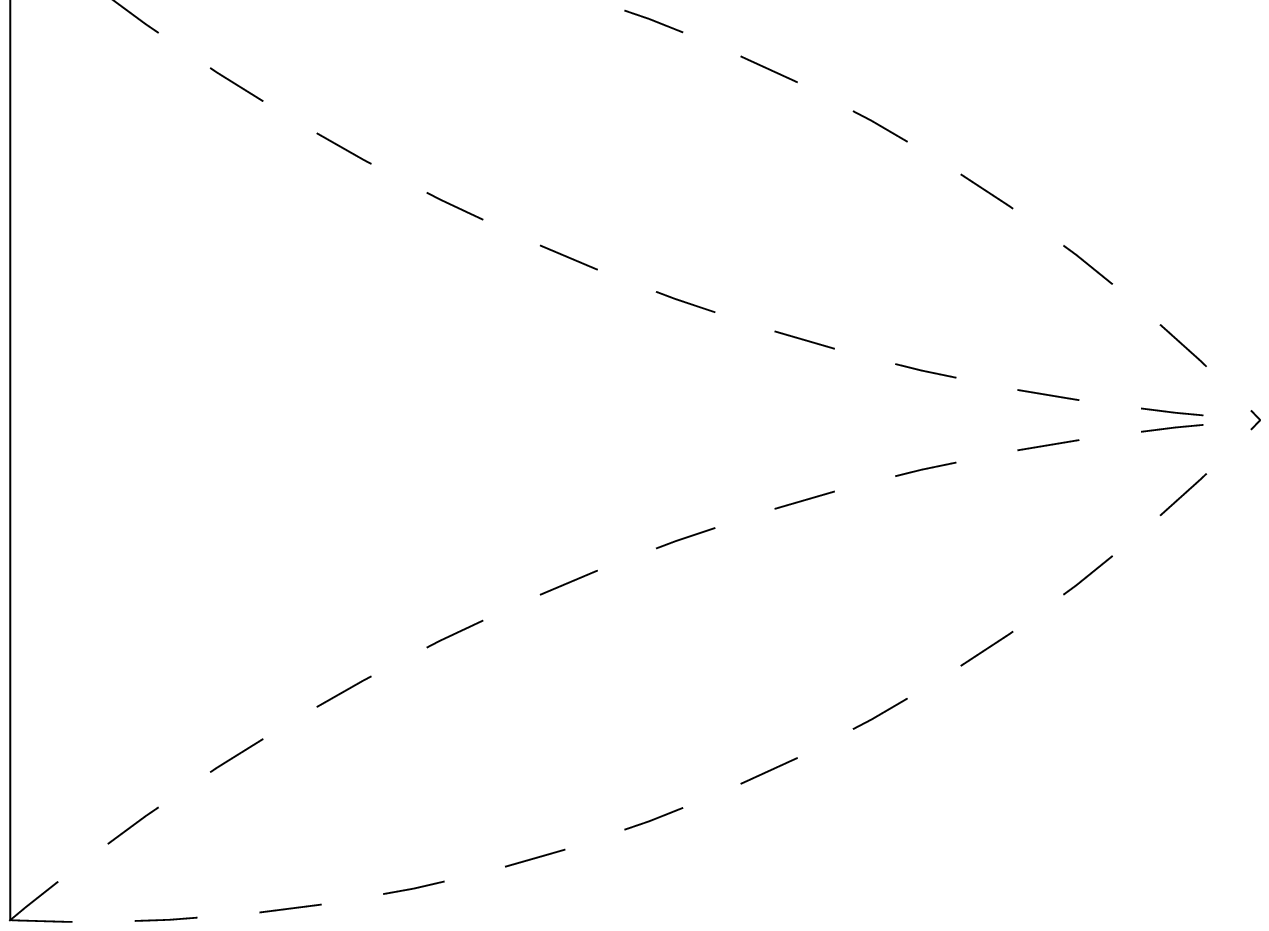}}  \hspace{-0.8cm}
+\frac{1}{16} \ \epsfxsize=2.6cm\epsfysize=3cm\raisebox{-2.4cm}{\epsffile{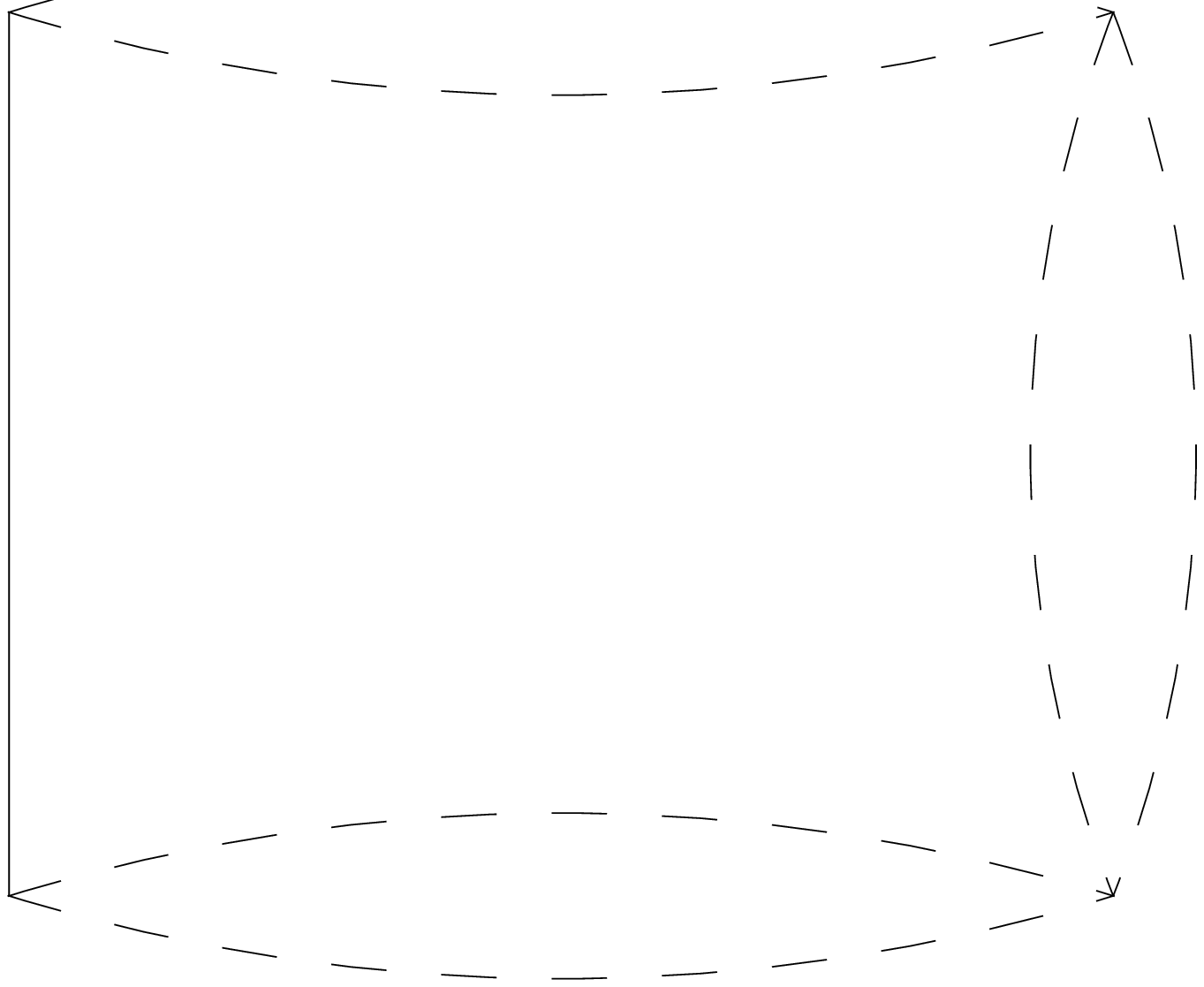}}  \hspace{-0.8cm}
+\dots\nn  \\[-0.9cm]&&
+\frac{1}{2} \ \epsfxsize=2.6cm\epsfysize=3cm\raisebox{-2.4cm}{\epsffile{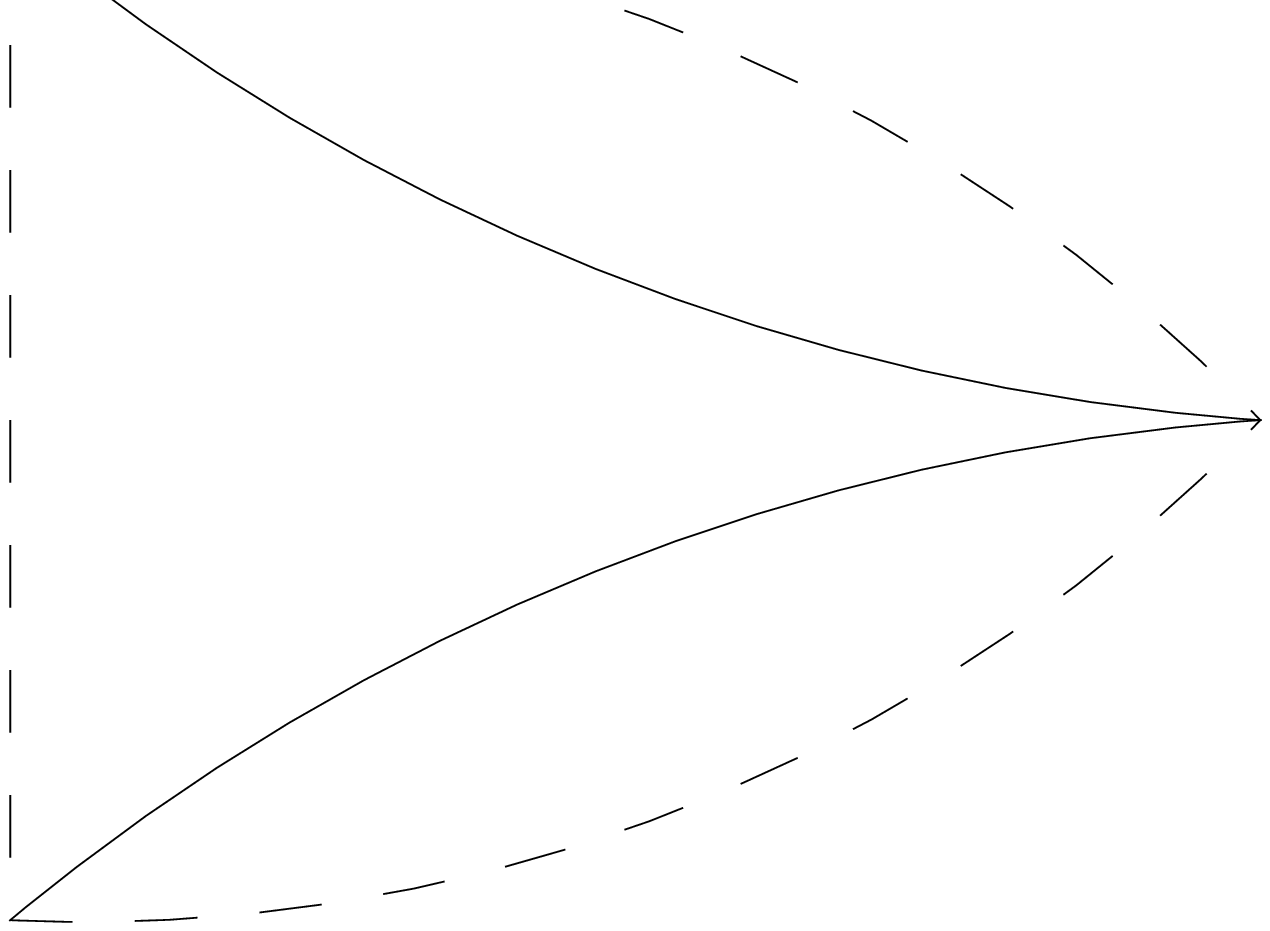}}  \hspace{-0.8cm} 
+\frac{1}{2} \ \epsfxsize=2.6cm\epsfysize=3cm\raisebox{-2.4cm}{\epsffile{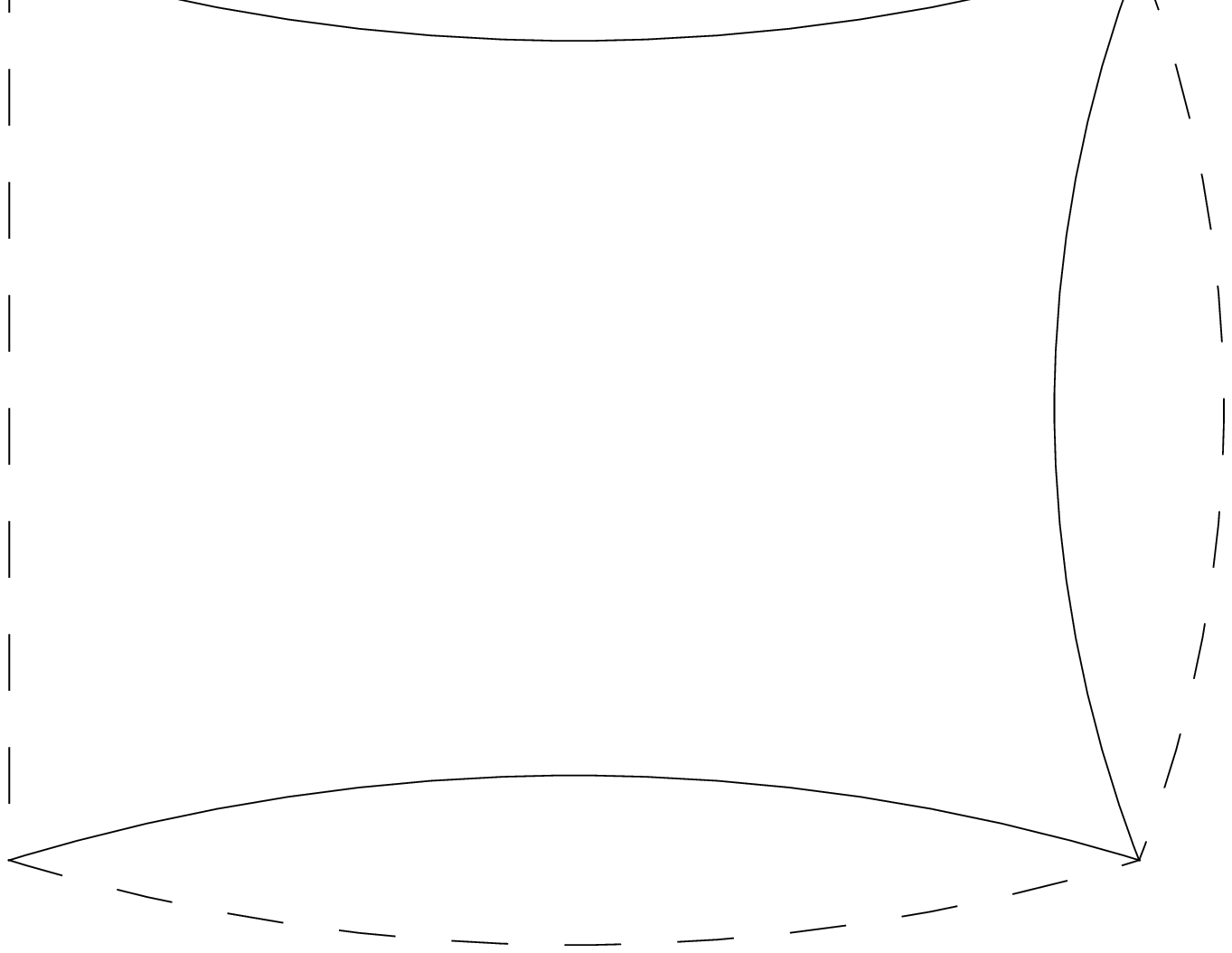}}  \hspace{-0.8cm} 
+\dots \ .\\[-0.9cm] \nn
\eea
These are all graphs forming rings (necklaces) of doubled lines (or
bubble chains). We denote them  by
\bea
\label{S1e}
\Sigma^{(1)}_\eta(p)&=&\Tr_q\beta_\eta(p+q)\beta_\eta(q) ,\nn \\
\Sigma^{(1)}_\phi(p)&=&\Tr_q\beta_\phi(p+q)\beta_\phi(q) ,  \nn \\
\Sigma^{(1)}_{\eta\phi}(p)&=&\Tr_q\beta_\eta(p+q)\beta_\phi(q) .
\eea
The loop integration in the graphs \Ref{Gamma} are all in the just
defined functions $\Sigma^{(1)}$ except for one integration flowing
through the single line. It is useful to introduce the following notations,
\bea\label{ab}
a&\equiv&\frac{-3\la}{N}\Sigma^{(1)}_\eta(p), \nn \\
b&\equiv&-\la\left(1+\frac{1}{N}\right)\Sigma^{(1)}_\phi(p), \nn \\
c&\equiv&\frac{-2\la}{N}\Sigma^{(1)}_{\eta\phi}(p) .
\eea
Here the vertex factors and the symmetry factor $(1/2)$ for
$\Sigma^{(1)}_\eta(p)$ and $\Sigma^{(1)}_\phi$ are included as well as
the factors resulting from the summation over the internal indices
$a,b,\dots =1,2,\dots,N-1$. For them we used
\[
 \sum \delta_{ab} \delta_{cd}=N-1 \equiv V_1 ,
\]
and
\[
\sum 
\delta_{ab}V_{aba_1b_1}V_{a_1b_1a_2b_2}\dots V_{a_{n-1}b_{n-1}cd}\delta_{cd}
=(N-1)\left(\frac{N+1}{3}\right)^{n-1}\equiv V_n .
\]
In terms of these abbreviations, $\Gamma$ can be written as
\bea\label{Gamma1}
\Gamma &=&\frac{-6\la}{N} \ \Tr_p\beta_\eta(p) \ \gamma_a 
-\frac{\la(N-1)}{N} \Tr_p \beta_\phi(p) \ \gamma_b  
\eea
with
\bea \label{gamma1}
\gamma_a &=& \frac{a}{6}+\frac12 a^2+\dots  +\frac12\ep b+\frac12\ep b^2 +\dots
 +\ep ab+\ep a^2b+\ep ab^2+\dots \  \\
&& +\frac12 c^2+\frac12 c^3 +\dots \ , \nn \\\label{gamma2}
\gamma_b&=&c^2+c^3+\dots ,
\eea
where the notation $\ep\equiv \frac{N-1}{3(N+1)}$ has been introduced
which appears as a factor in front of each sequence of symbols $b$.

The symmetry factors of the graphs result in factors $\frac12$ for
sequences of symbols which are symmetric under transposition. Non
symmetric sequences like $ab$ appear only once because $ba$ describes
the same graph. Now we pull out the symmetry factor $\frac12$ and
correct that for the non symmetric sequences by writing them twice,
e.g., $ab+ba$ instead of $ab$.

In this way we are faced with the sum over all sequences of symbols
$a$ and $b$. As a symbolic notation for the subgraphs in $\Gamma$ they
have to be taken to be non commuting. After inserting the analytical
expressions in momentum space representation, in fact they do commute.
But first let us sum up these sequences.

We observe that for each sequence containing a single factor $a$,
i.e. a fragment $\dots bab\dots$, there is a sequence containing
$baab$ instead, $baaab$, etc. These sequences can be summed up
partially,  
\[ \sum_{\i=1}^\infty \dots b\underbrace{a\dots a}_ib\dots =
\dots b Ab\dots \ ,
\]
where the notation
\be\label{A} A\equiv \frac{a}{1-a}
\ee
was introduced.  With other words, multiple neighbored factors $a$
appear after this partial resummation only within the combination
$A$. An exception is the sequence consisting of one single $a$ which
is treated separately by writing $\frac{a}{3}=a-\frac{2a}{3}$. The
same holds for the sequences of symbols $b$ where in addition a factor
$\ep$ has to be taken into account. The summation results in
\[ \sum_{\i=1}^\infty \dots a\ep\underbrace{b\dots b}_ia\dots =
\dots a Ba\dots
\]
with
\be\label{B}  B\equiv \frac{\ep b}{1-b} .
\ee
After these   resummations,$\gamma_a$, \Ref{gamma1}, takes the form
\[
\gamma_a=\frac12 \left( -\frac{2a}{3}+A+B+A^2+B^2+AB+BA+\dots \right) \ .
\]
Now the sequence of $A$'s and $B$'s can be summed up and we arrive
finally at
\be\label{gamma3}
\gamma_a=\frac12\left(-\frac{2a}{3}+\frac{A+B+2AB}{1-AB}  \right) .
\ee
The sequences containing $c$ form a simple geometric series 
\[
\sum_{i=2}^\infty c^i= cC \ 
\]
 and with the notation 
\be\label{C} C=\frac{c}{1-c} .
\ee
we obtain
\be\label{gamma4} \gamma_b=c \ C.
\ee
In this way, by means of Eqs.  \ref{gamma3} and \Ref{gamma4} we
obtained the complete expression for the functional $\Gamma$,
Eq. \Ref{Gamma1}.

Next we have to consider the functional $D$ appearing in $W_2$,
Eq. \Ref{W2app0}. It is the sum of all graphs containing only quartic
vertices and at least 2 loops.  Again, within the approximation made in
Sect. 4, we restrict ourselves to the contributions consisting of
rings of doubled lines. In the case where there is one field only the
corresponding graphical representation is
\bea\label{Done}
D&=&\frac{1}{48} \ \epsfxsize=2.6cm\epsfysize=3cm\raisebox{-2.0cm}{\epsffile{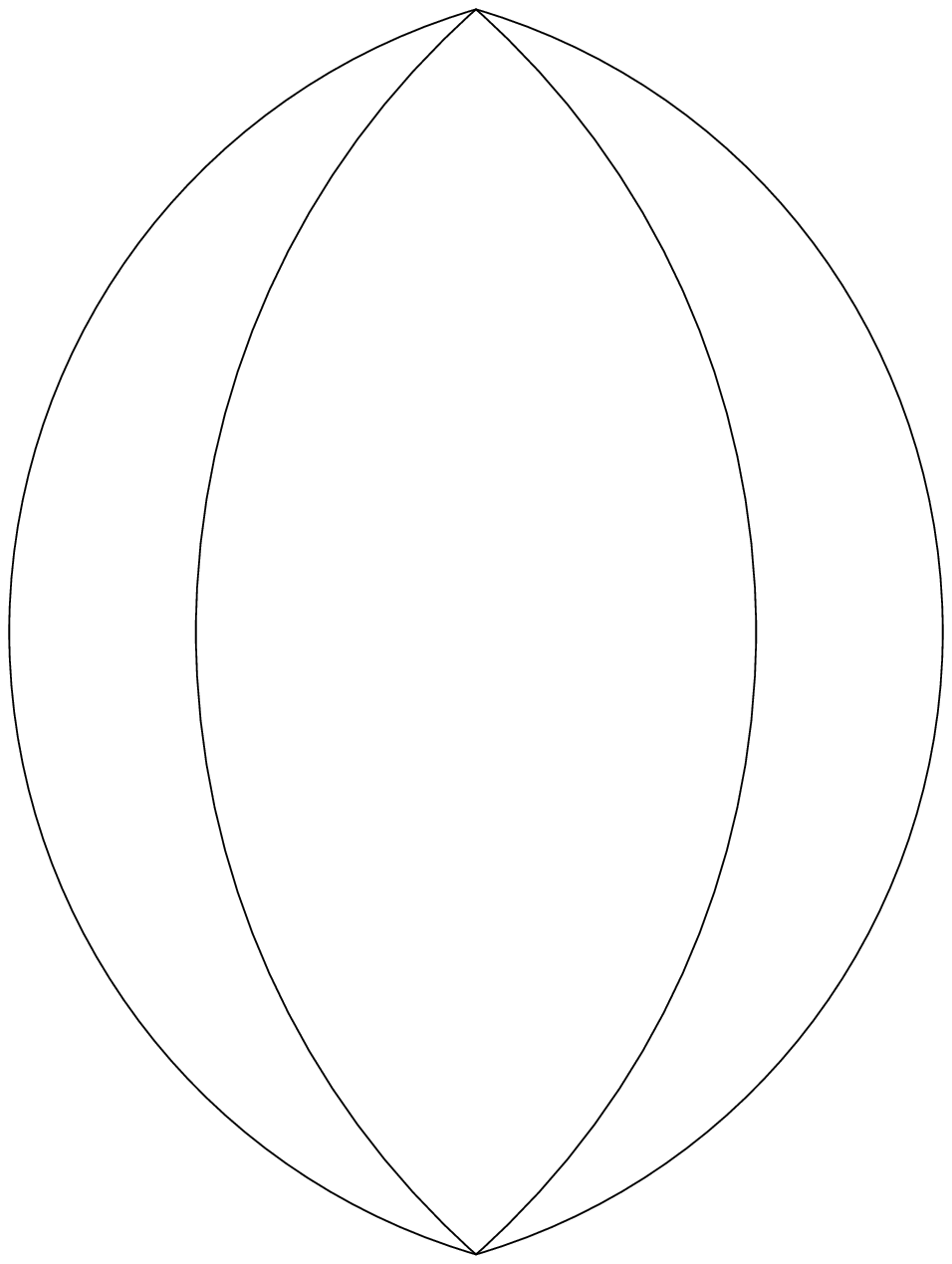}} \hspace{-1.3cm} 
+\frac{1}{48}   \  \epsfxsize=2.6cm\epsfysize=3cm\raisebox{-2.0cm}{\epsffile{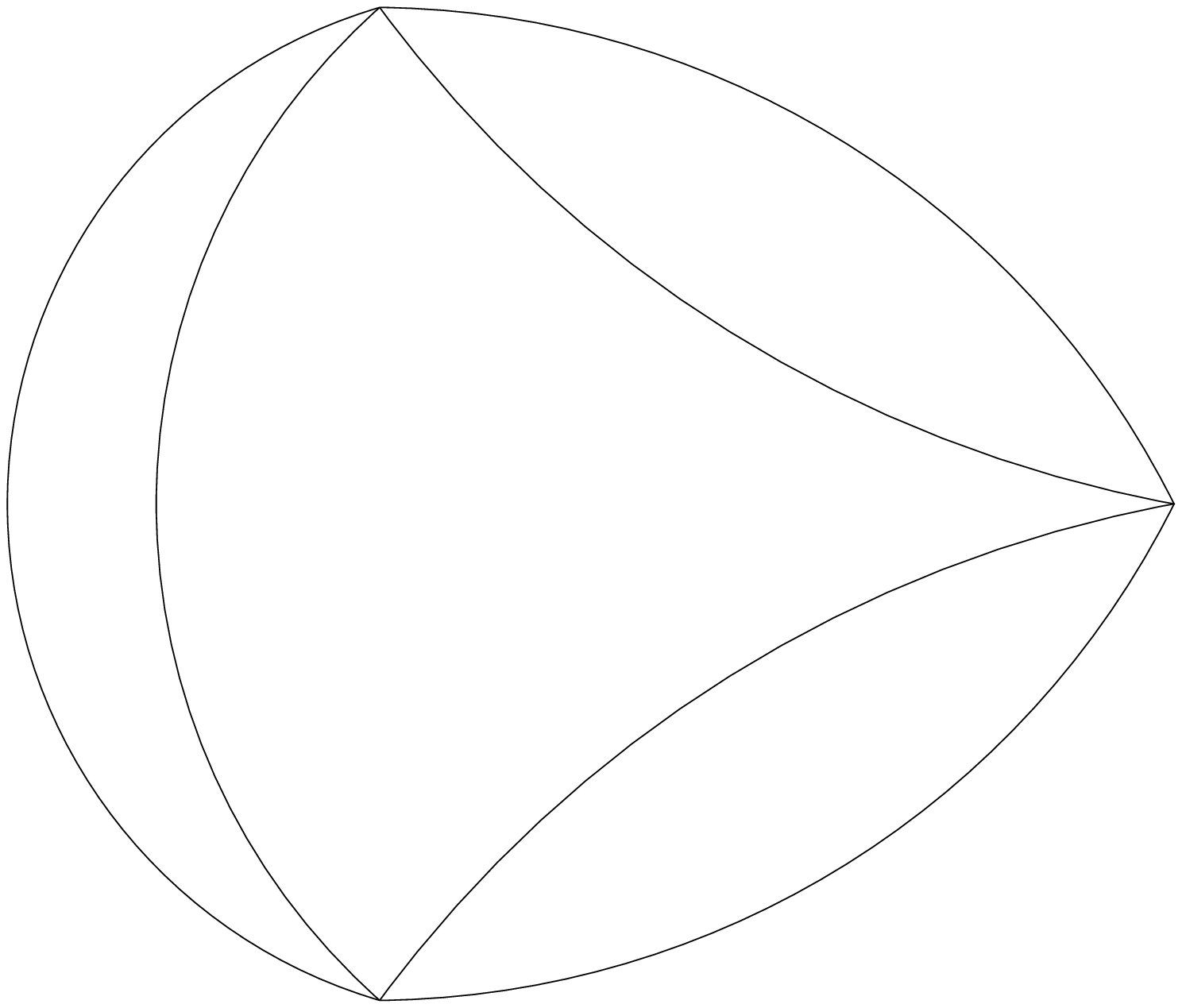}} \hspace{-0.6cm} 
+\frac{1}{128}   \ \epsfxsize=2.6cm\epsfysize=3cm\raisebox{-2.0cm}{\epsffile{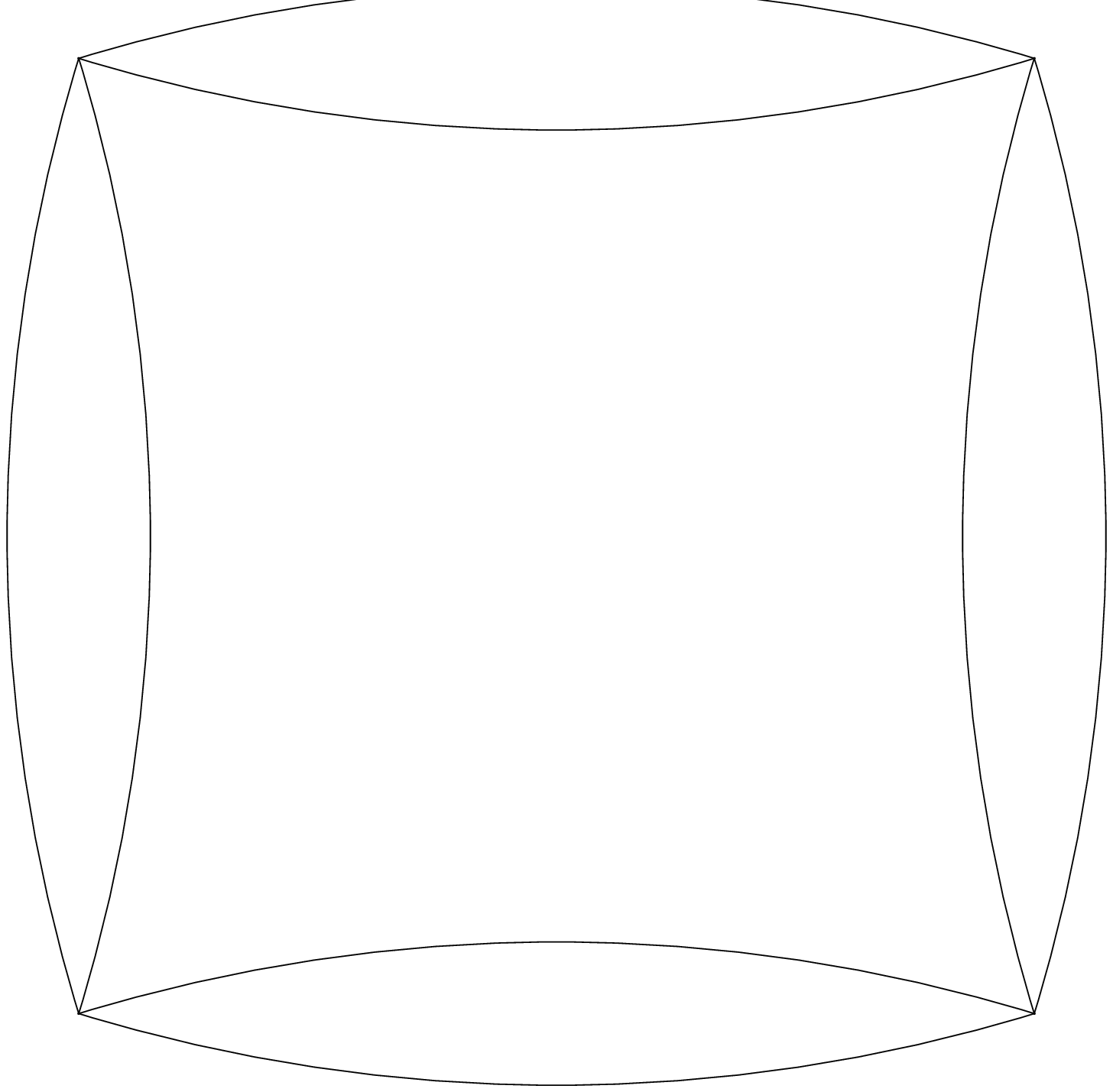}} \hspace{-0.1cm} 
+                   \dots \ , \nn \\
&=& 
\frac{1}{48}  \ \epsfxsize=2.6cm\epsfysize=3cm\raisebox{-2.0cm}{\epsffile{d2.ps}} \hspace{-1.2cm} 
+\sum_{n\ge3}^\infty \frac{1}{  2^n \ n} \epsfxsize=2.6cm\epsfysize=3cm\raisebox{-2.0cm}{\epsffile{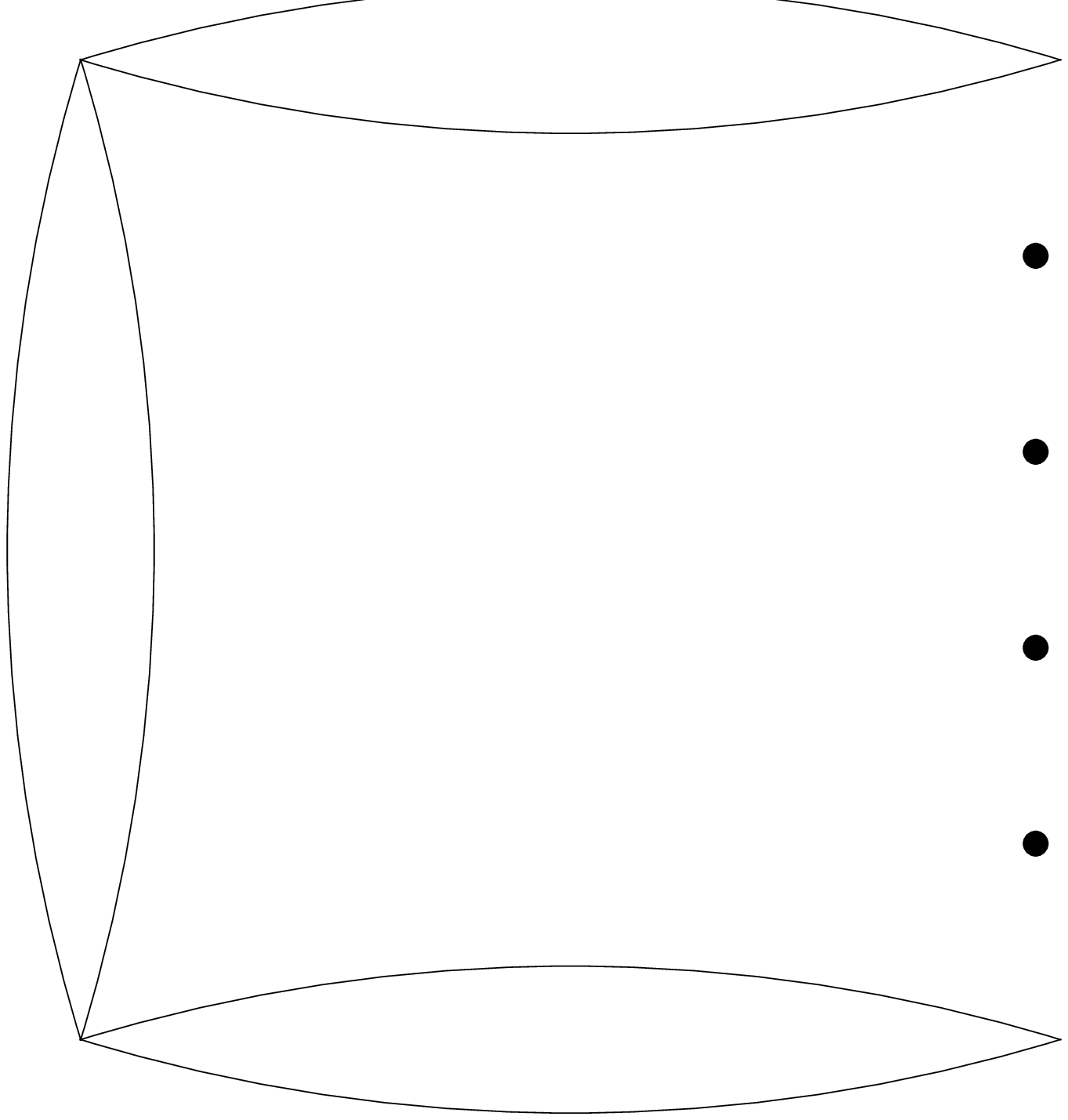}} \, ,
\hspace{-0.8cm}\\[-1.3cm] \nn
\eea 
which is the same as Eq. (9) in \cite{Bordag:2000tb}.  In the $N$-component case
with unbroken symmetry the corresponding expression are essentially
the same. For broken symmetry they become more complicated.  However,
in the gap equations we need the derivatives $\delta D/\delta
\beta_\eta(p)$ and $\delta D/\delta \beta_\phi(p)$ only. By means of the relation
\[ \Gamma = \frac{\delta D}{\delta \beta_\eta(p)}
\]
one of them is already calculated and we are left with $\delta
D/\delta \beta_\phi(p)$.  The graphs appearing here are quite similar
to that in $\Gamma$, Eq. \Ref{Gamma}, where now the single line is
$\beta_\phi$ instead of $\beta_\eta$.
The graphical representation of a generic graph is
\bea\label{Dgen}
\frac{\delta D}{\delta (\beta_\phi(p))_{ab}}&=&  \epsfxsize=2.6cm\epsfysize=3cm\raisebox{-2.0cm}{\epsffile{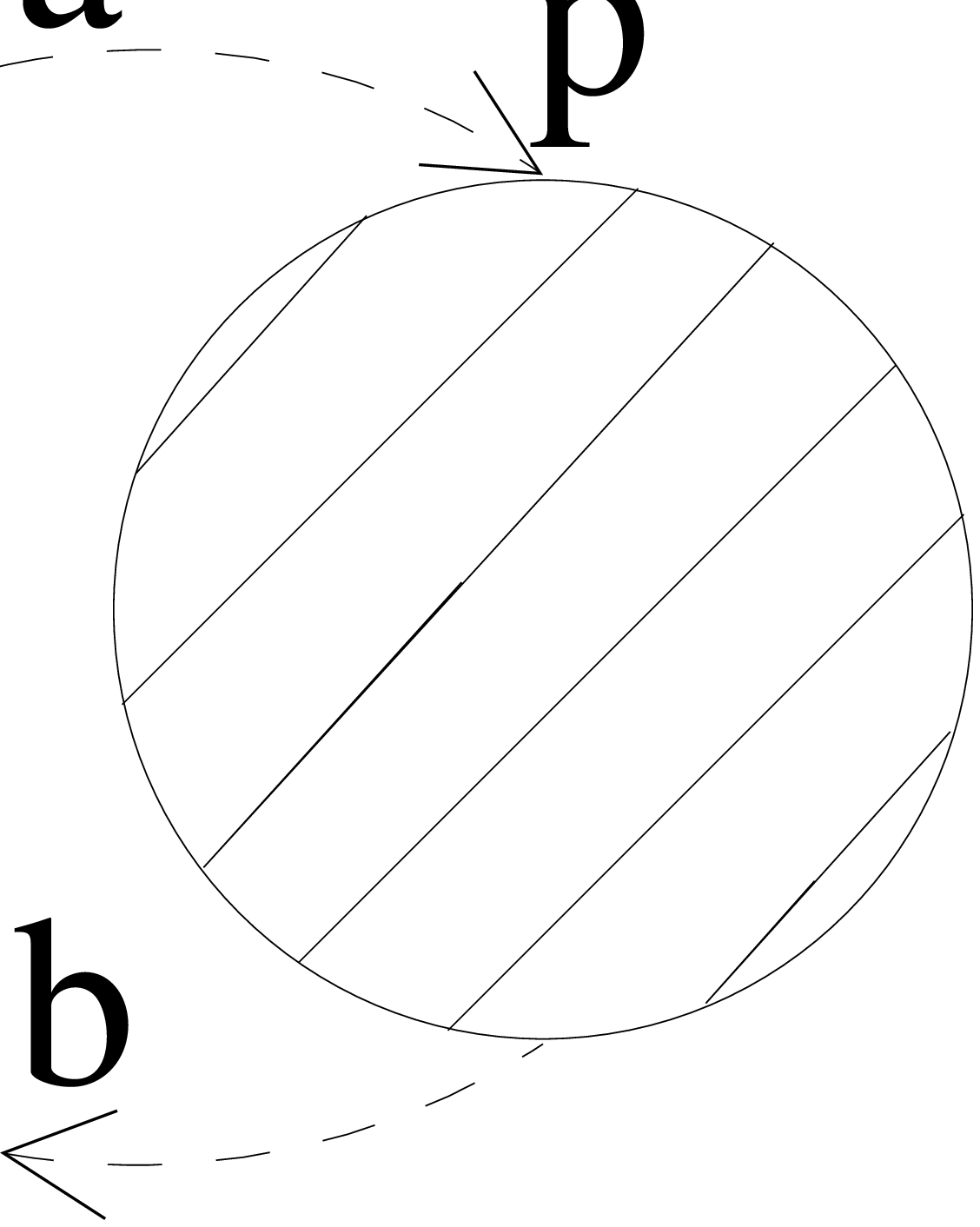}}
 \ , \\[-1.3cm] \nn\eea
where $p$ is the momentum on the external line and $a$ and $b$ are the
indices on the lines. Theses indices are to summed up according to
\[ \frac{\delta D}{\delta \beta_\phi(p)}= 
\sum_{a,b=1}^{N-1}\delta_{ab}\frac{\delta D}{\delta (\beta_\phi(p))_{ab}} \, .
\]
The first few graphs read then
\bea\label{Dgen1}
\frac{\delta D}{\delta \beta_\phi(p)}&=& 
\frac{1}{12} \ \epsfxsize=2.6cm\epsfysize=2.4cm\raisebox{-1.7cm}{\epsffile{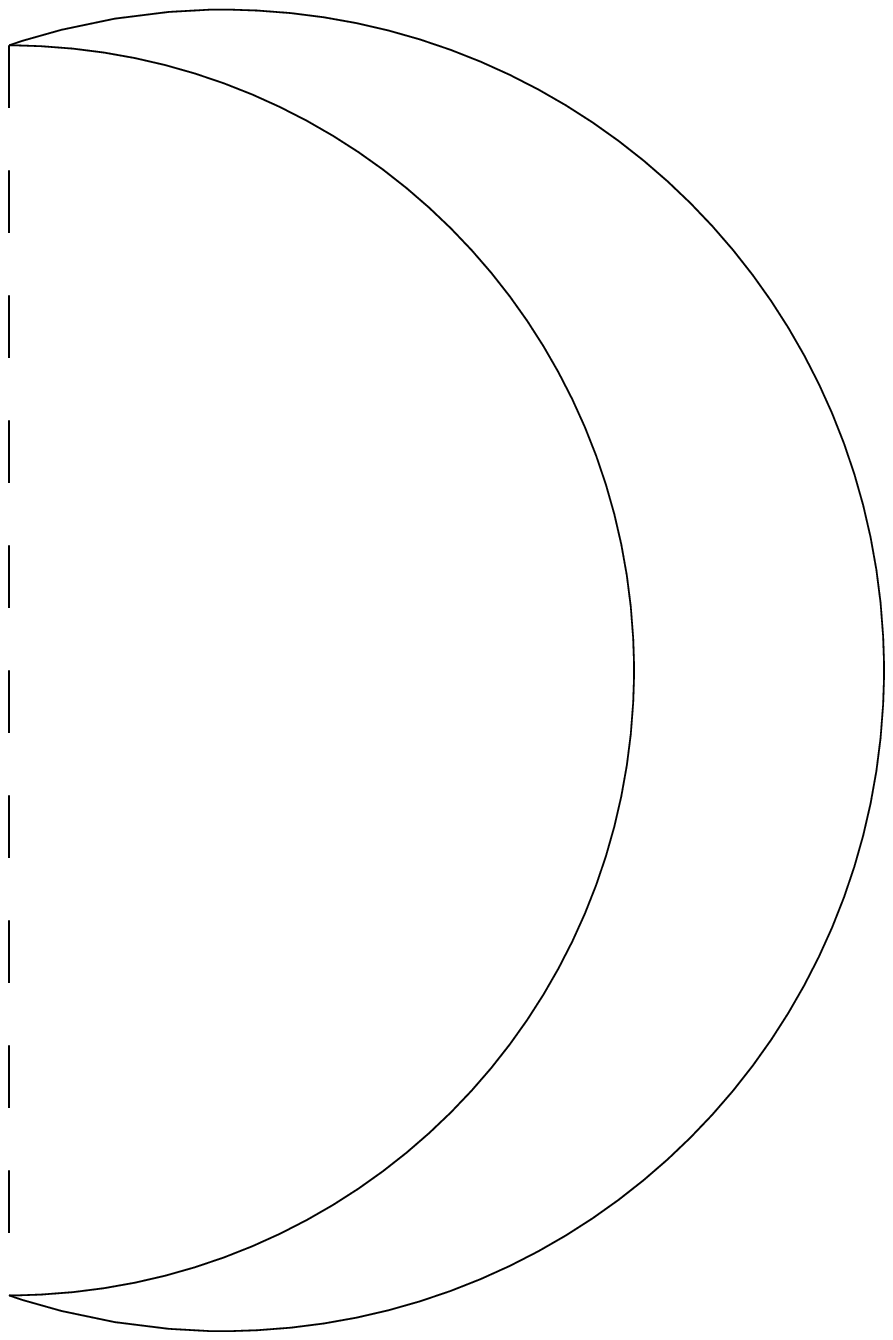}} \hspace{-0.8cm}
+\frac{1}{8} \ \epsfxsize=2.6cm\epsfysize=3cm\raisebox{-2.4cm}{\epsffile{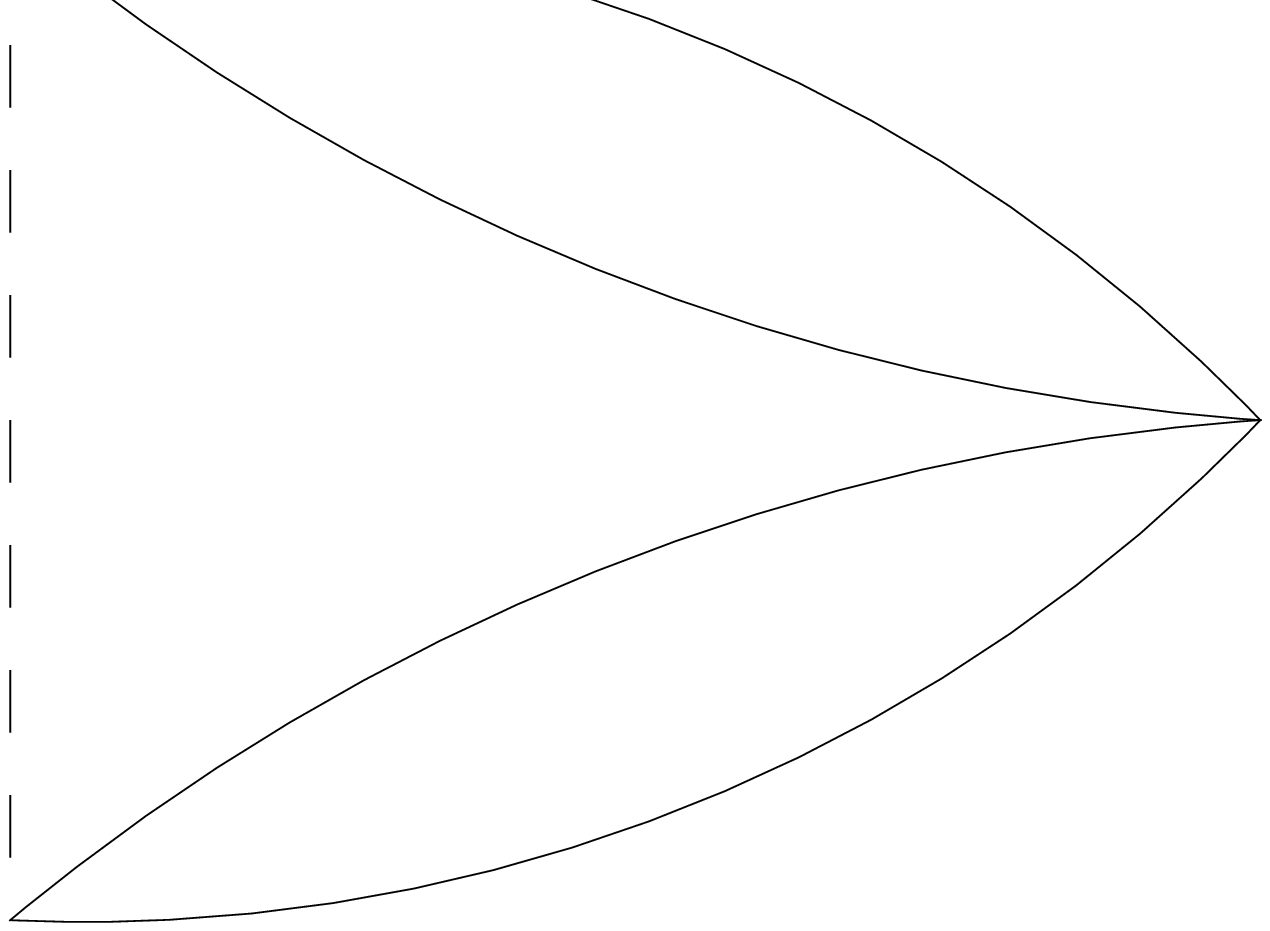}}  \hspace{-0.8cm}
+\frac{1}{16} \ \epsfxsize=2.6cm\epsfysize=3cm\raisebox{-2.4cm}{\epsffile{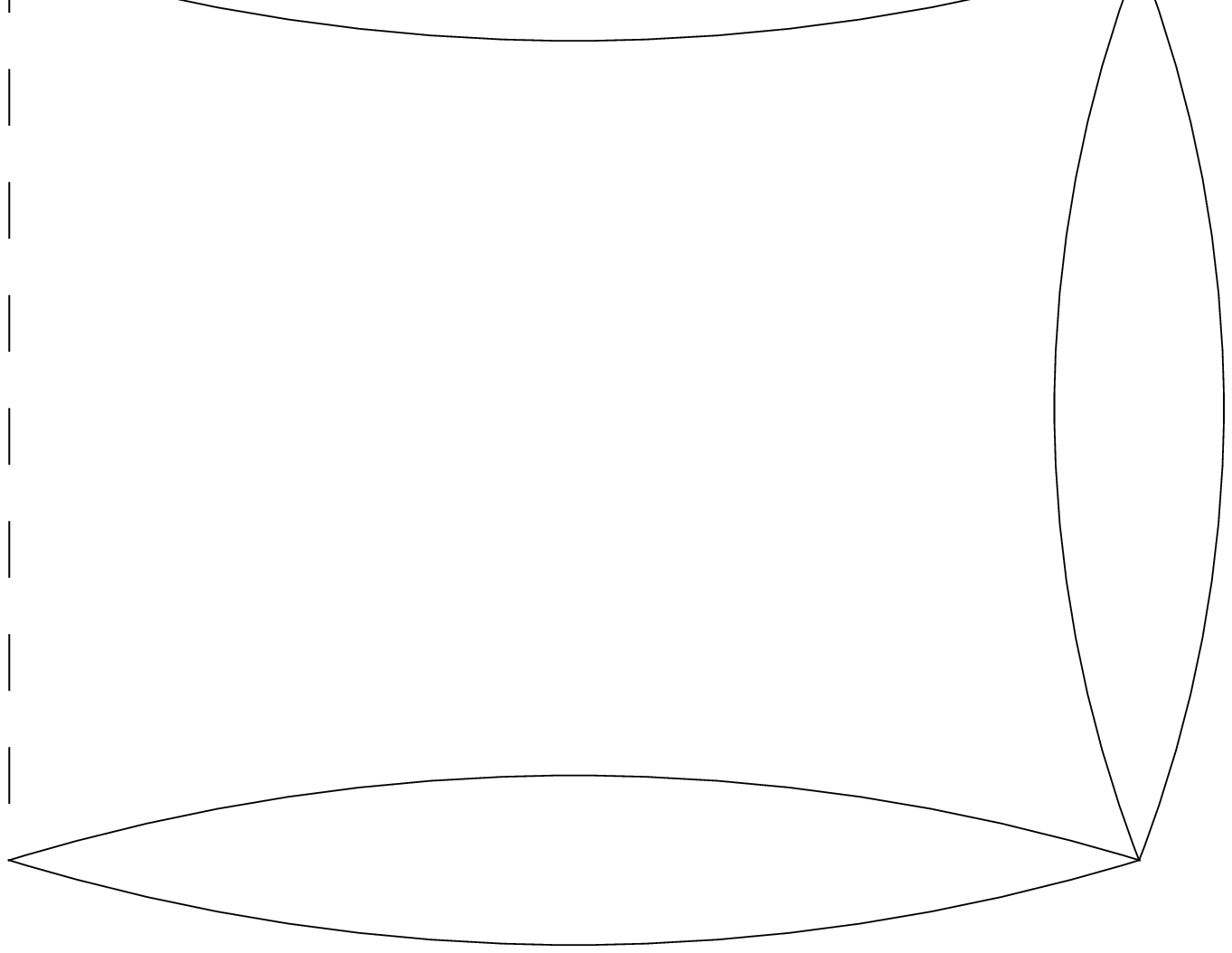}}  \hspace{-0.8cm}
+\dots\nn \\[-0.9cm]&&
+\frac{1}{4} \ \epsfxsize=2.6cm\epsfysize=3cm\raisebox{-2.4cm}{\epsffile{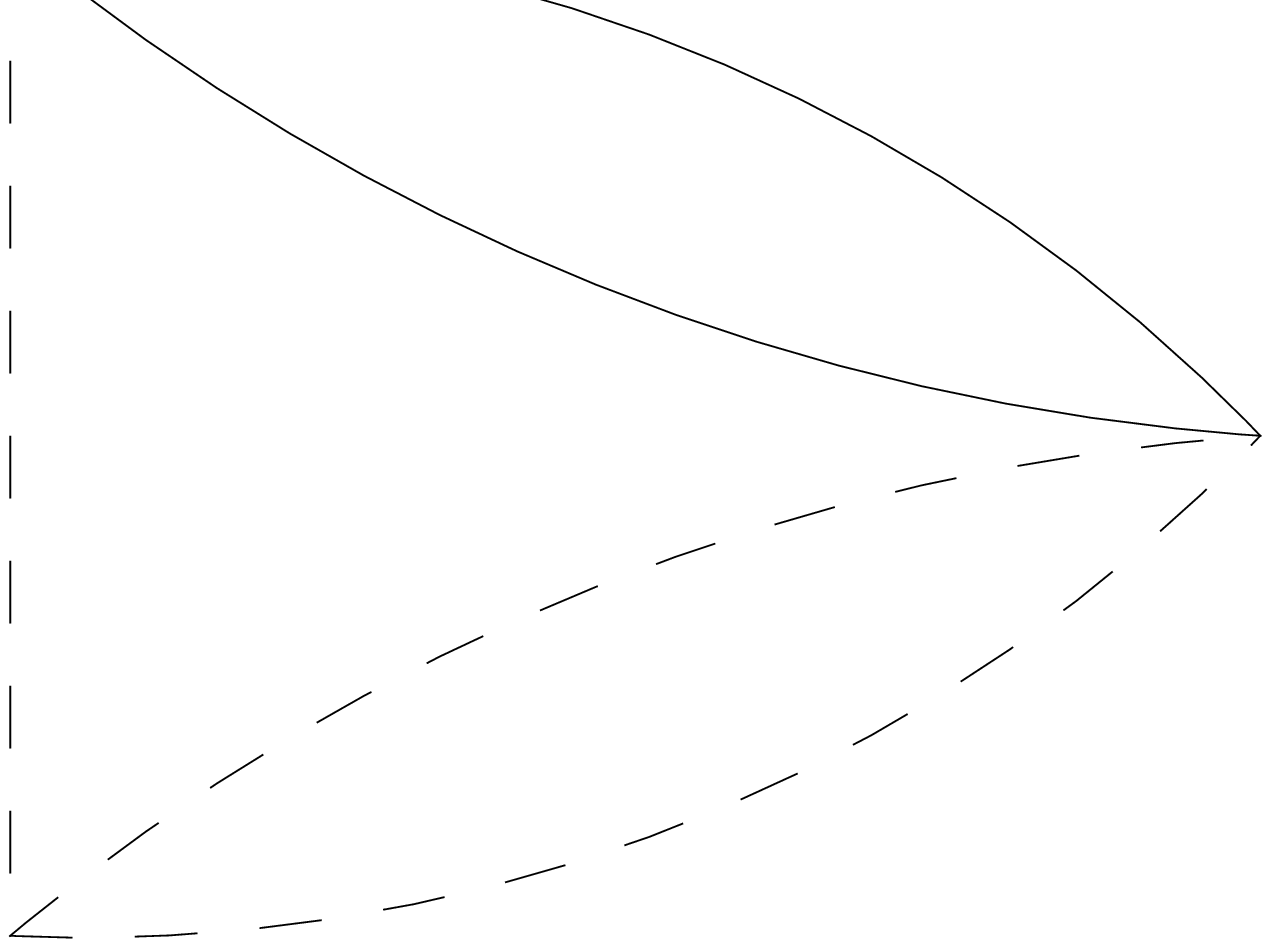}}  \hspace{-0.8cm}
+\frac{1}{8} \ \epsfxsize=2.6cm\epsfysize=3cm\raisebox{-2.4cm}{\epsffile{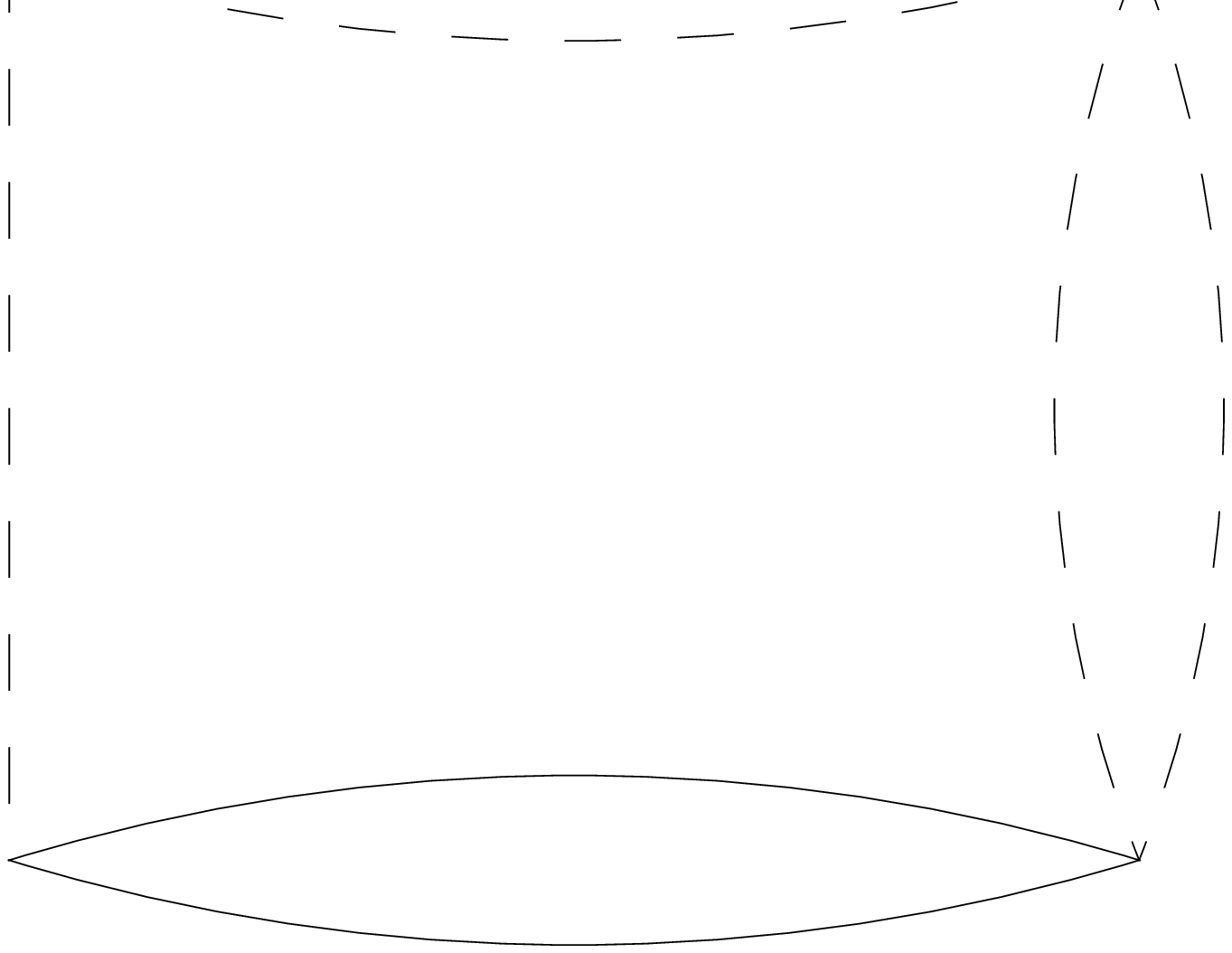}}  \hspace{-0.8cm}
+\frac{1}{8} \ \epsfxsize=2.6cm\epsfysize=3cm\raisebox{-2.4cm}{\epsffile{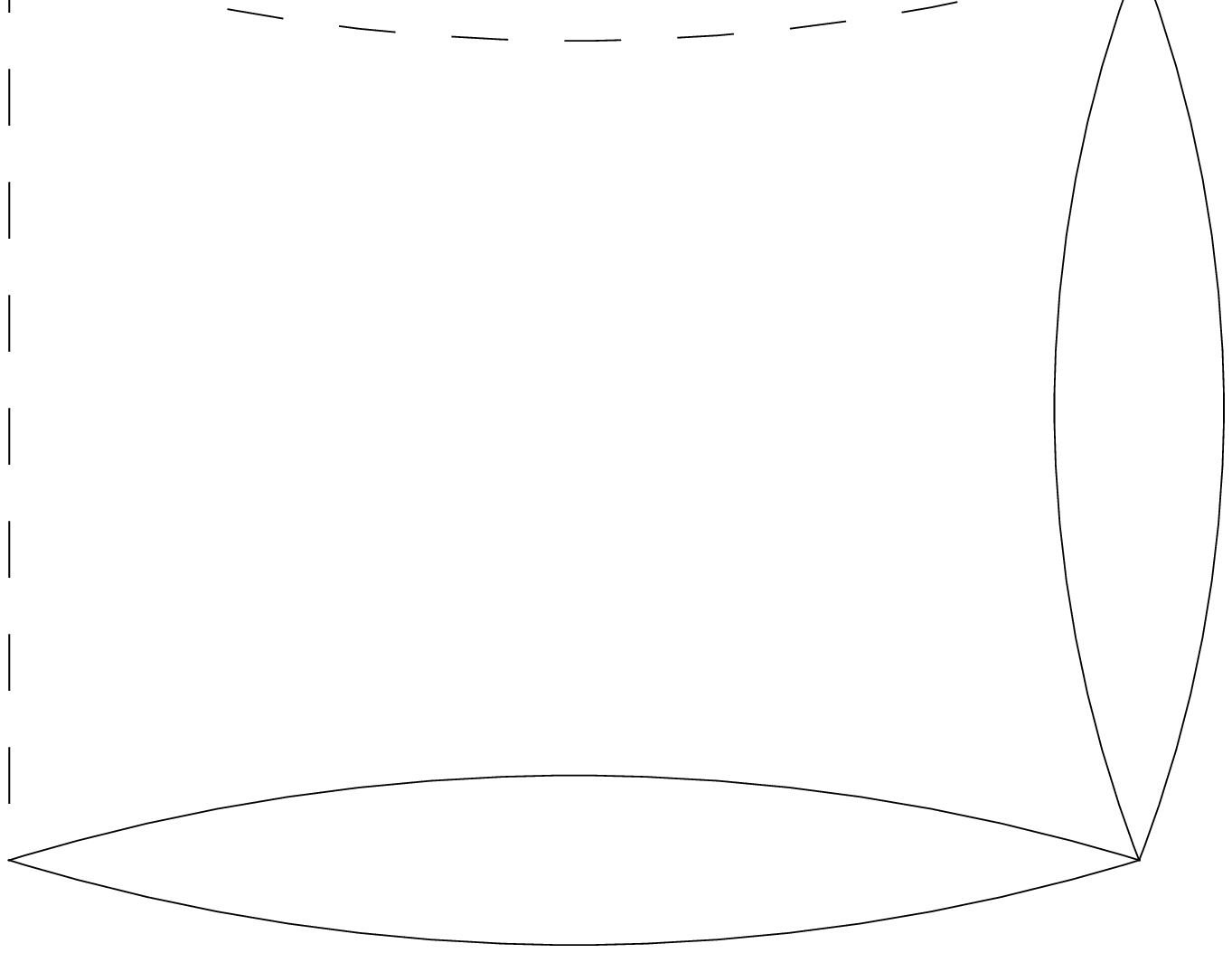}}  \hspace{-0.8cm}
+\dots\nn  \\[-0.9cm]&&
+\frac{1}{4} \ \epsfxsize=2.6cm\epsfysize=2.4cm\raisebox{-1.7cm}{\epsffile{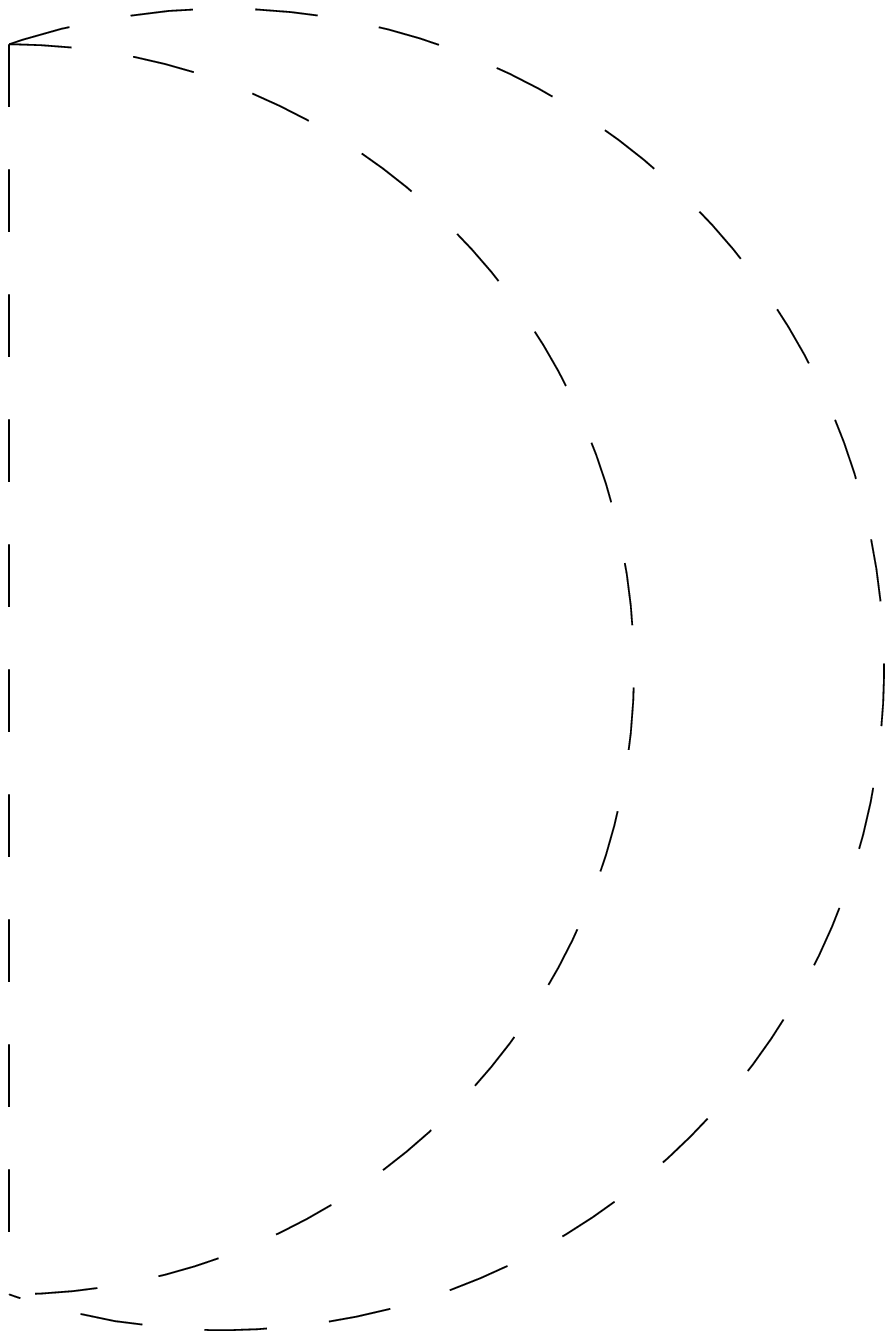}}  \hspace{-0.8cm}
+\frac{1}{8} \ \epsfxsize=2.6cm\epsfysize=3cm\raisebox{-2.4cm}{\epsffile{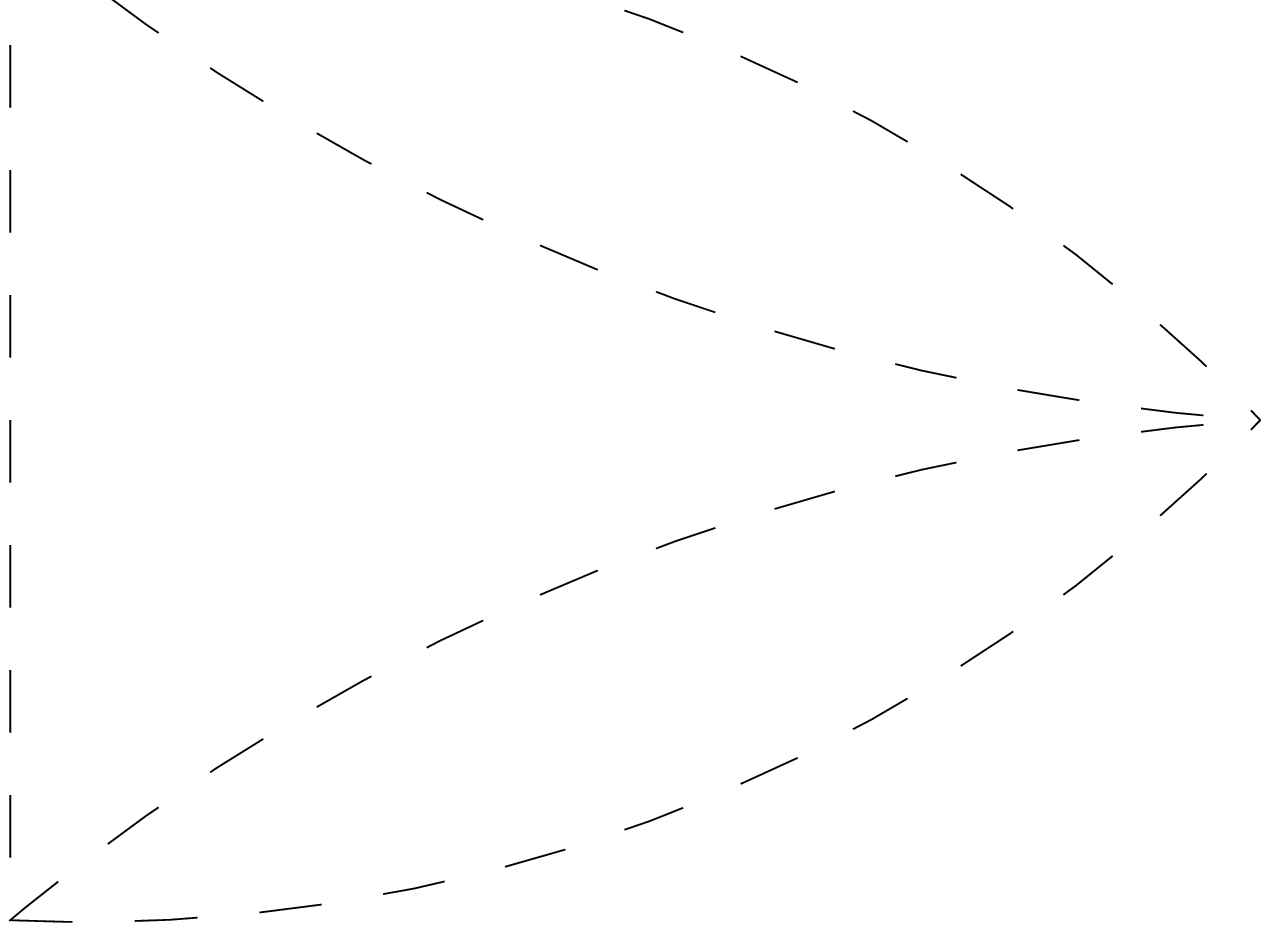}}  \hspace{-0.8cm}
+\frac{1}{16} \ \epsfxsize=2.6cm\epsfysize=3cm\raisebox{-2.3cm}{\epsffile{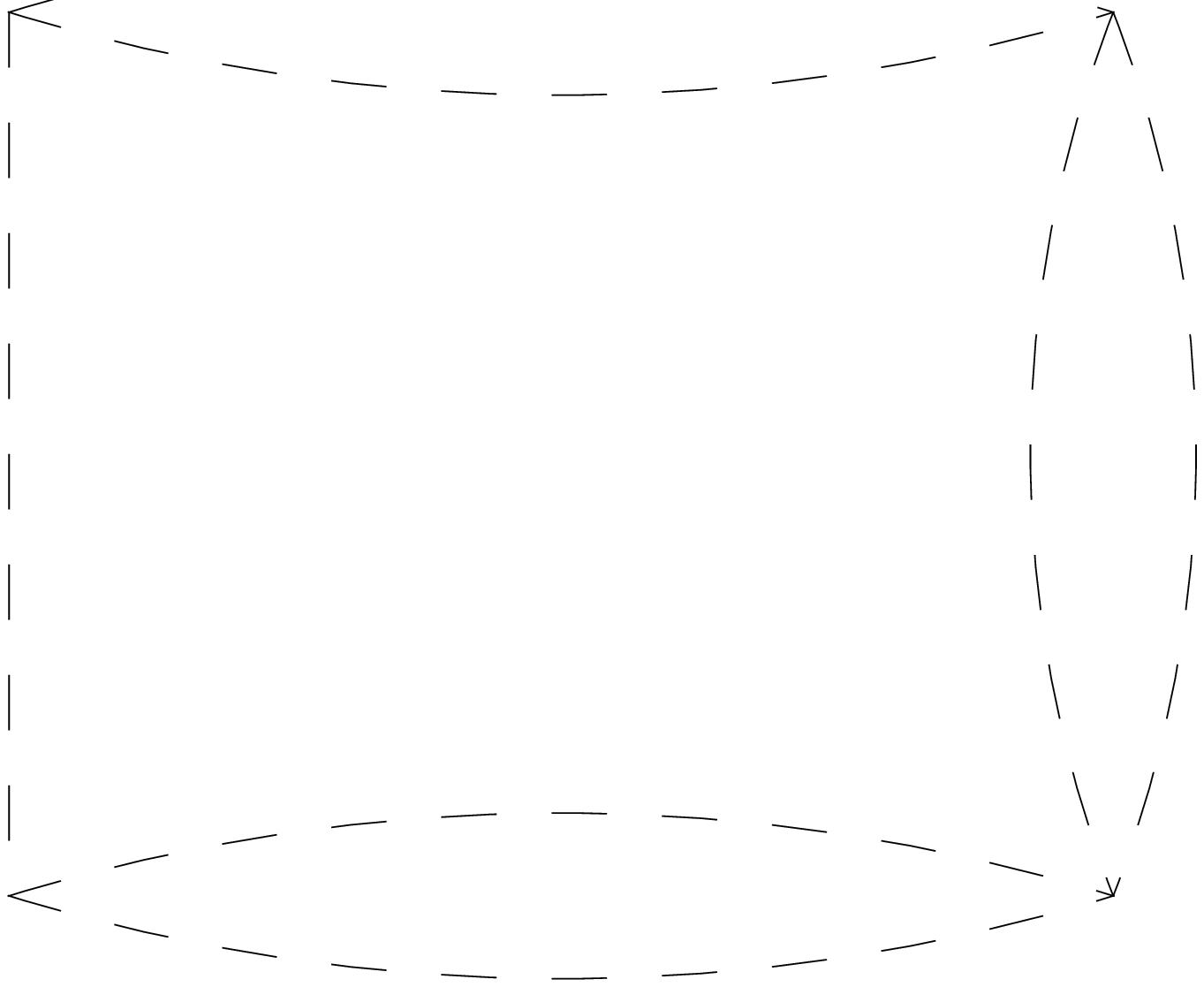}}  \hspace{-0.8cm}
+\dots\nn  \\[-0.9cm]&&
+\frac{1}{2} \ \epsfxsize=2.6cm\epsfysize=3cm\raisebox{-2.4cm}{\epsffile{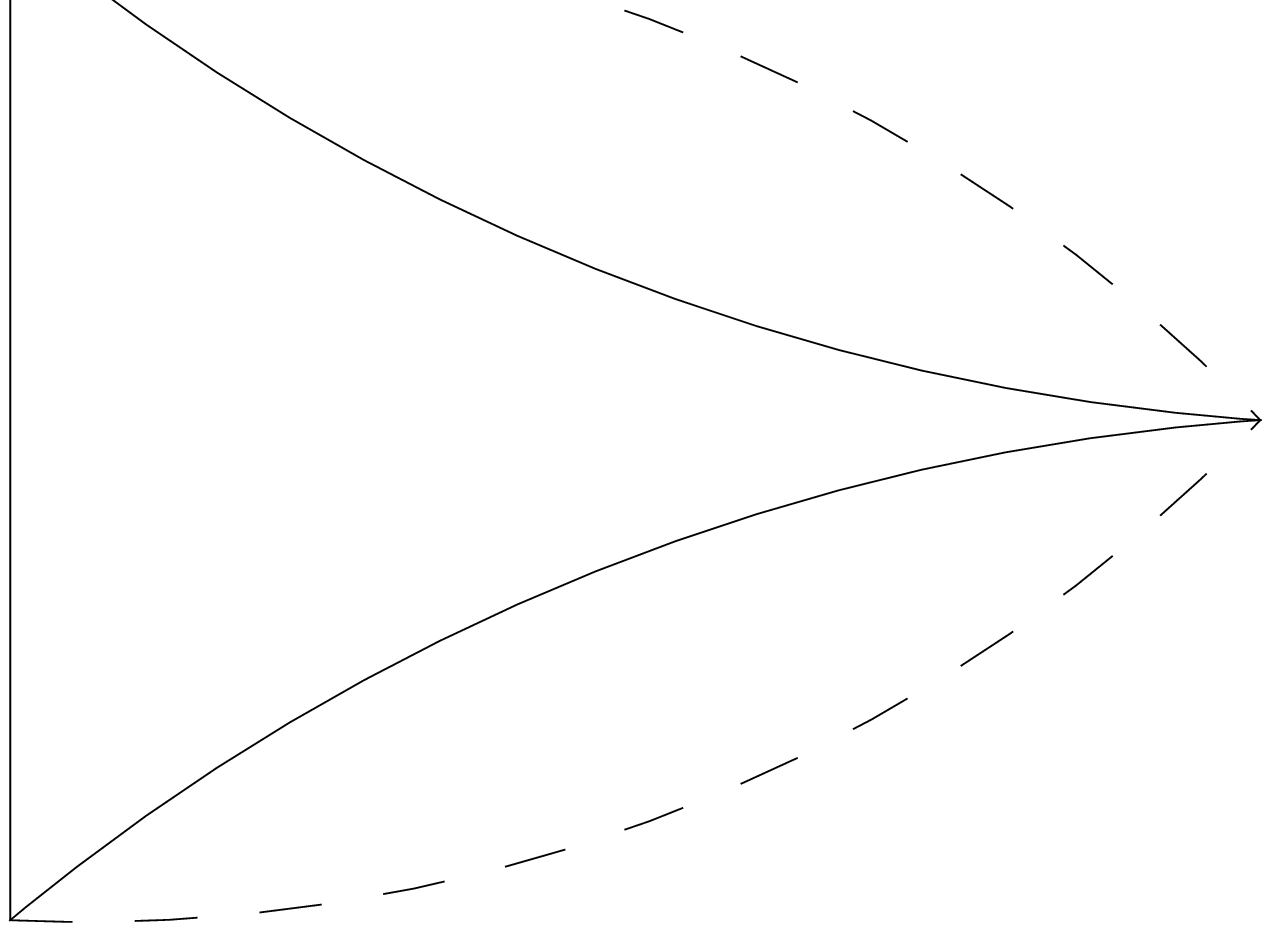}}  \hspace{-0.8cm} 
+\frac{1}{2} \ \epsfxsize=2.6cm\epsfysize=3cm\raisebox{-2.4cm}{\epsffile{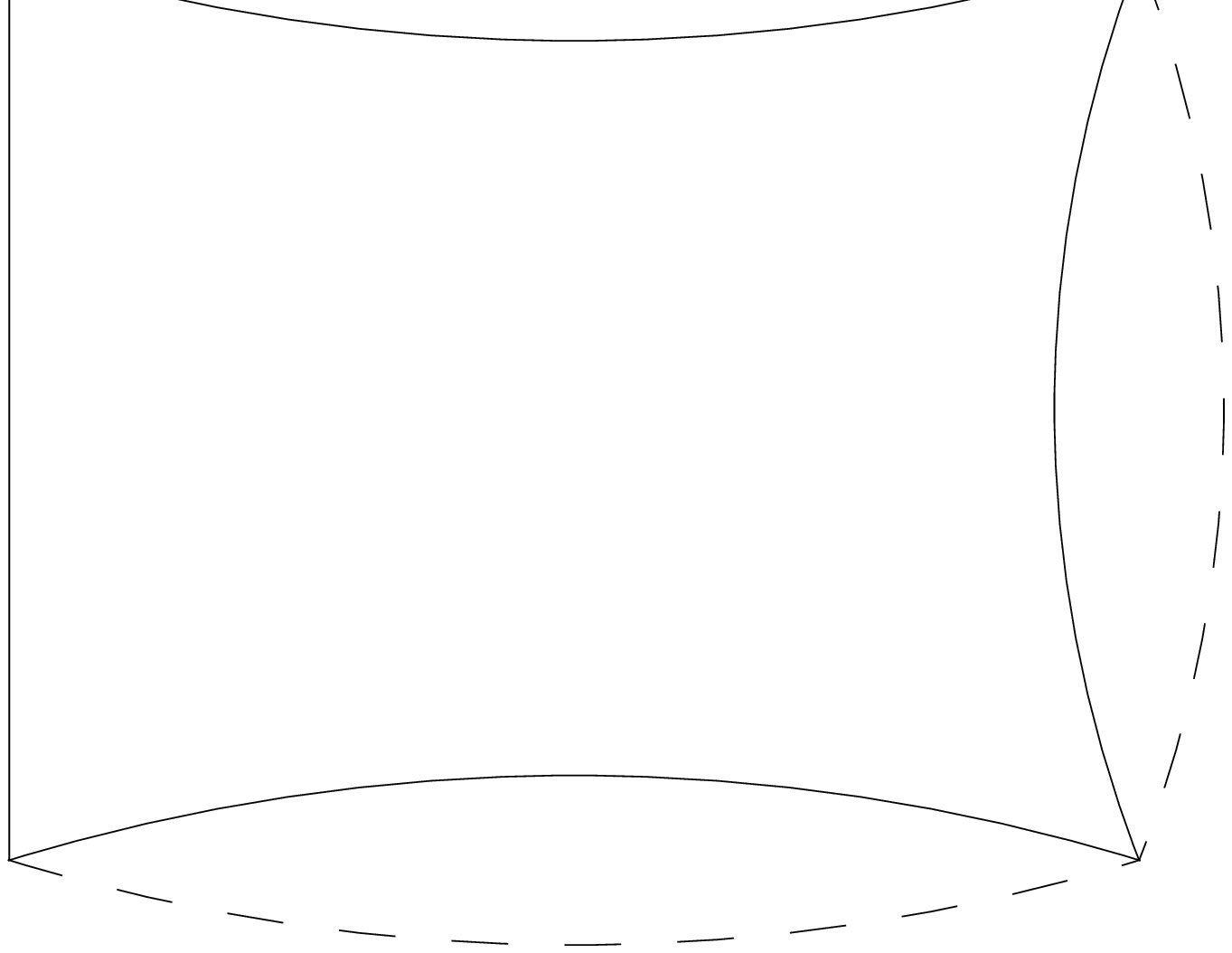}}  \hspace{-0.8cm} 
+\dots \ .\\[-1.2cm] \nn\eea
We write down the corresponding analytical expression in the form
\bea\label{dD} 
 \frac{\delta D}{\delta \beta_\phi(p)}&=&
   \Tr_q \beta_\phi(p+q) \ (\gamma_1+\gamma_2+\gamma_3) \\
&& +\Tr_q \beta_\eta(p+q) \ \gamma_4 , \nn
\eea
where the $\gamma_i$'s correspond to the raws in Eq. \Ref{Dgen1}. 

For $\gamma_1$ we have
\[
\gamma_1=
\frac{1}{12}\frac{-6\la}{N} \Sigma^{(1)}_\phi(q) \V_1 +
\sum_{i\ge2}\frac{1}{2^{n+1}}\left(\frac{-6\la}{N}\right)^{n+1} 
\left(\Sigma^{(1)}_\phi(q)\right)^n \V_n \ ,
\]
where the $\V_n$ result from the summation over the internal indices according to
\beao
\V_1&\equiv&\sum V_{aba_1b_1}V_{a_1b_1ab}=\frac{N^2-1}{3} ,\\
\V_n&\equiv&\sum V_{aba_1b_1}V_{a_1b_1a_2b_2}\dots V_{a_nb_nab}=
\frac{N+1}{3}\left[\left(\frac{N+1}{3}\right)^n+\left(\frac{2}{3}\right)^n(N-2)\right].
\eeao
With the notations $a$, $b$ and $c$ given by Eq. \Ref{ab} and
$\tilde{b}=\frac{2}{N+1}b$ we sum up the geometric series and arive at
\be\label{g1}\gamma_1=
\frac{-\la(N+1)}{2N} \left[ \frac{b}{3}
+\frac{b^2}{1-b}+(N+2)\left(\frac{\tilde{b}}3+\frac{\tilde{b}}{1-\tilde{b}}\right)\right].
\ee
With the remaining graphs we proceed in the same manner. For $\gamma_2$ we note
\bea\label{g2}\gamma_2&=& 
\frac14 \frac{-6\la}{N}  \left(\frac{-2\la}{N}\right)^2 
\Sigma^{(1)}_\eta(q)\Sigma^{(1)}_\phi(q) V_1 \nn \\
&&+\frac18 \left(\frac{-6\la}{N}\right)^2 \left(\frac{-2\la}{N}\right)^2
\Tr_q  \left(\left(\Sigma^{(1)}_\eta(q)\right)^2\Sigma^{(1)}_\phi(q) V_2
+\Sigma^{(1)}_\eta(q)\left(\Sigma^{(1)}_\phi(q)\right)^2V_1\right) +\dots \nn \\
&=& \frac{-3\la}{N} \frac{N-1}{9} \left(\frac{1}{\ep}ab+\frac{1}{\ep}a^2b+\frac{1}{\ep}ab^2+\dots \right),
\eea
which is essential the same as $\gamma_a$ in Eq. \Ref{gamma1} except for the dependence on $\ep$. Summing up we obtain
\be\label{gamma2a} 
\gamma_2=\frac{-3\la}{N} \frac{N-1}{9}\left(a+aA+\frac{1}{\ep}\frac{\left(2+\ep A+\frac{1}{\ep}B\right)AB}{1-AB}\right).
\ee
For $\gamma_3$ we have simply
\be\label{gamma3a}\gamma_3=\frac{-\la}{N}(N-1)  \ cC
\ee
so that taking all together we arrive at
\bea\label{Df}\frac{\delta D}{\delta\beta_\phi(p)}&=&
-\frac{\la}{N}\Bigg\{\Tr_q\beta_\phi(p+q)\Bigg[  (N+1)\left[ \frac{b}{3}+\frac{b^2}{1-b}+(N-2)\left(\frac{\tilde{b}}{3}+\frac{\tilde{b}^2}{1-\tilde{b}}\right)\right] \nn \\
&& ~~~~~~~~~~~~~~~~~~~~~~~~~~(N-1)\left[ \frac{a}{3}+\frac 13 \ aA+\frac{1}{3\ep}\frac{\left(2+A+\frac{1}{\ep}B\right)AB}{1-AB}\right] \Bigg] \nn\\
&&~~~~~+\Tr_q\beta_\eta(p+q) (N-1) \frac{c^2}{1-c}  \Bigg\},
\eea
where $\tilde{b}=\frac{2}{N+1}b$.


\end{document}